\documentclass[final,5p,times,twocolumn]{elsarticle}

\usepackage{amssymb, amsfonts, amsmath, bm, amstext}
\everymath{\displaystyle}
\usepackage{subfigure}
\usepackage{hyperref}
\usepackage{amssymb,amsfonts,amsthm}
\usepackage{geometry}
\usepackage{graphics,subfigure}
\usepackage{hyperref}

\usepackage{color}
 \usepackage{natbib}
 \biboptions{sort&compress}
\begin{document}

\begin{frontmatter}

\title{A novel buckling pattern in periodically porous elastomers with applications to elastic wave regulations}

\author[add1,add2]{Yang Liu}
\author[add1]{Tian Liang}
\author[add1]{Yuxin Fu}
\author[add1,add2]{{Yu-Xin Xie}\corref{cor1}}
\ead{xyx@tju.edu.cn}
\author[add1,add2]{Yue-Sheng Wang}
\address[add1]{Department of Mechanics, School of Mechanical Engineering, Tianjin University, Tianjin 300354, China}
\address[add2]{Tianjin Key Laboratory of Modern Engineering Mechanics, Tianjin 300354, China}
\cortext[cor1]{Corresponding author.}

\begin{abstract}
This paper proposes a new metamaterial structure consisting of a periodically porous elastomer with pore coatings. This design enables us to engender finite deformation by a contactless load. As a case study, we apply thermal load to the pore coating and carry out a finite element analysis to probe instabilities and the associated phononic properties. It turns out that a novel buckling mode, preserving the nature of surface wrinkling in tubular structures, can be induced under a plane-strain setup, and a smaller size of the unit cell is attained compared to the counterpart of traditional buckled profile in soft porous elastomers. In particular, this buckling pattern is able to produce several bandgaps in different frequency ranges as the macroscopic mean strain increases. We further introduce a metallic core as local resonator, and the updated metamaterial allows a low-frequency bandgap, the bandgap width of which can be estimated by a simplified theoretical model. As more free parameters are involved in the structure, we perform a detailed parametric study to elucidate the influences of the modulus ratio between coating and matrix, the porosity, the core radius, and the macroscopic mean strain on the buckling initiation and the evolution of bandgap. Remarkably, a stiffer surface coating is prone to enhance the stability of the structure, which is contrary to existing results in film/substrate bilayers. It is expected that the current study could shed light on new insight into pattern formation and wave manipulation in porous elastomers. 
\end{abstract}


\end{frontmatter}
%%%%%%%%%%%%%%%%%%%%%%%%%%%%%%%%%%%%%%%%%%%%%%%%%%%%%%%%%%%%%%%%%%%%
\section{Introduction}
Phononic crystals, composed of periodical materials or structures, are a category of metamaterials and have unfolded great potential in multiple engineering applications in recent years due to their promising abilities for regulating elastic, electromagnetic, or acoustic waves. In general, bandgaps wherein these waves of specific frequencies are able to be prevented may be generated in phononic crystals by Bragg scattering \citep{khd1993, hhs2006} or local resonance \citep{lzm2000,zlh2012}, etc. This prominent feature promotes the use of phononic crystals as a prospective medium in wave filters \citep{pdv2004,qlm2005,czm2017}, waveguides \citep{bkc2005,cdb2012,wwl2018}, noise suppressions \citep{gcs2012,xww2012,zwx2015}, vibration reductions \citep{aso2012,cbc2012,xwy2013}, and they hence attract much research interest in the past decades. More recent advances and prospects of phononic crystals can be found in the monographs by  Deymier \citep{deymier2013} and Laude \citep{laude2015} and the review article by Wang et al. \cite{www2020}

In principle, the frequency range of a bandgap can be tuned by proper designs of the physical properties and geometrical shapes of phononic crystals. Recently, periodically porous elastomers have soon become a research hotspot as they usually experience large elastic deformation in response to various external stimuli. Furthermore, various buckling patterns may be created as well \citep{mdb2007,bbd2008}. This paves an effective way to reconfigure material structures and further to bring a remarkable change of wave properties.  Bertoldi and Boyce \cite{bb2008,bb2008w} found that a particular buckling pattern in periodically porous structures produces bandgaps and then allows to forbid the propagation of waves within some frequency ranges. Since then, diverse studies were dedicated to understanding the influences of the loading scenarios, material nonlinearities, pore arrangements and geometries, porosities, pattern transitions, etc. on the pattern formation and the accompanied bandgaps in soft metamaterials.  Overvelde et al. \cite{osb2012} discussed the role of pore shape in controlling the compaction performance through buckling based on both experimental and numerical results and clarified that the circular hole is not an optimal shape. Wang et al. \cite{wsb2013} revealed the effects of geometrical and material nonlinearities, as well as the applied strain on the evolution of bandgaps by considering an equi-biaxial compression and by employing the neo-Hookean and Gent material constitutions. It is found that the tunability of bandgaps can be enhanced by amplifying the geometrical nonlinearity. Note that the circular pores are squarely arrayed in the above-mentioned works. In order to increase the tunability, non-circular holes, multiple arrangements, or local resonators can be introduced.  Shan et al. \cite{skw2014} exploited various folding modes in a periodic elastomer with a triangular array of pores to regulate elastic waves. They discovered that different loading directions result in different buckling patterns and further lead to multiple dynamical behaviors of the metamaterial. In addition, it is also demonstrated that rhombille and deltoidal trihexagonal arrangements of pores can be applied to altering the positions and widths of bandgaps \citep{swb2015}. Also, the case of criss-crossed elliptical holes was examined by  Gao et al.\cite{ghb2018, glb2019} where additional bandgaps can be generated compared to the counterparts of circular holes. Besides, an alternative way for manipulating wave propagations was paved by arranging local resonators \citep{wcs2014}, inclusions \citep{lwc2019}, or hard scatters \citep{nly2020,nyl2020} inside the porous elastomers. It should be pointed out that there is plentiful literature in this field, especially in recent years, and not all of which have been surveyed here. Interested readers can refer to the review article by  Wang et al. \cite{www2020}.

It is worth mentioning that a specific instability type is triggered by compression or stretch in the above-mentioned studies. As indicated earlier, other loading methods can be applied to practical situations such as swelling \citep{jkb2009} and deflation \citep{cj2018}. In particular, swelling can be regarded as a prototypical non-contact loading approach. Jang et al.  \cite{jkb2009} experimentally reported a pattern transformation in nanoscale periodic structures made of SU8 material induced by solvent swelling and evaporation and explored the associated changes in the phononic band structure. Zhang et al. \cite{zcy2019} studied bandgaps in soft network metamaterials comprised of sandwiched horseshoe microstructures created by unusual swelling behaviors. Chen and Jin \cite{cj2018} designed a new metamaterial with non-uniform sized holes where geometrical instability can be incurred by deflation and presented a reduced beam model to identify the bifurcation load. Later, a refined model was proposed by Liang et al. \cite{lfl2021} for predicting the critical pressure considering elliptical pores and the dependence of surface texture modulation on deflation-induced buckling pattern was examined by Fernandes et al. \cite{fmf2022}. Generally speaking, a deformed configuration will recover its initial state once the external loading is removed. Therefore, many researchers utilized the glass transition temperature in polymers to freeze a buckling pattern \citep{ymd2018,knr2020,hrw2021}. For the dynamical response, Konarski et al. \cite{knr2020} first heated the fabricated porous elastomer over its glass transition temperature and then generated a desired deformation by axial compression which can be fixed by a rapid cooling effect. The ability to control underwater scattering was inspected as well. Also, Hu et al. \cite{hrw2021} engineered a lattice metamaterial and explored the bandgaps at distinct temperatures. 

It is known that the mechanism behind bandgaps in soft porous elastomers is different from that occurring in traditional hard materials, although the cavity geometry is also a vital parameter for bandgaps in hard porous metamaterials. For instance, Li et al. \cite{lnl2020} carried out an inverse design of the hole shapes via a deep learning algorithm towards an expected bandgap frequency. They unfolded that a wider bandgap can be generated if all holes occupy a profile with more wrinkles. On the one hand, wrinkled morphologies are ubiquitous in nature, such as intestines \citep{bc2013} and pumpkins \citep{dl2014}. These patterns are created in the growth process as a consequence of solution bifurcation. In our systematic investigations for growing tubular tissues, the scaling laws for the critical buckling load and critical wavenumber as well as the amplitude equation have been derived \citep{jlc2018,jlc2019}. Meanwhile, the effects of material and growth gradients on pattern formation were demonstrated \citep{lzdc2020,llc2021}. On the other hand, surface wrinkling in porous elastomers has not been addressed as far as we know. The main reason is that there is usually one category of soft material in periodically porous elastomers but surface wrinkling can only be triggered in a bilayer system. As a result, it is well-motivated to produce such a typical buckling mode in porous elastomers by coating a circular layer in each cavity. Despite the complexity of fabrication, it definitely increases the possibility of regulating the wave propagations. Furthermore, differing from the methodology that produces deformation by compression or stretch, we shall explore a contactless loading approach, for instance, swelling, heating, or magnetic field. Meanwhile, as a primary exploration, we focus on a finite element study of instability and wave propagation using commercial software Abaqus \citep{abaqus}. 

This paper is organized as follows. We illustrate the metamaterial structure and the associated unit cell in Section 2. A finite element model is established and the corresponding nonlinear analysis is performed in Section 3. In particular, the unit cell is found to be a $1\times1$ structure where only an intact hole is involved for the proposed metamaterial structure subject to heating. We investigate the dynamical response of the designed metamaterial in Section 4 and identify several narrow bandgaps in the high-frequency ranges. To induce a possible bandgap in low-frequency ranges, we make use of a metallic core embedded in the center of each hole, and the corresponding static and dynamic responses are unraveled in Section 5. A simplified theoretical model is constructed as well to evaluate the upper and lower bounds of the low-frequency bandgap. Section 6 exhibits an exhaustive parametric study on both the critical buckling load and the evolution of bandgaps as functions of the modulus ratio between coating and matrix, the porosity, the applied mean macroscopic strain, and the core radius. The paper is briefly summarized in Section 7.

\section{Metamaterial structure and modeling}
        
\begin{figure*}[!ht]
\centering\includegraphics[width=15cm]{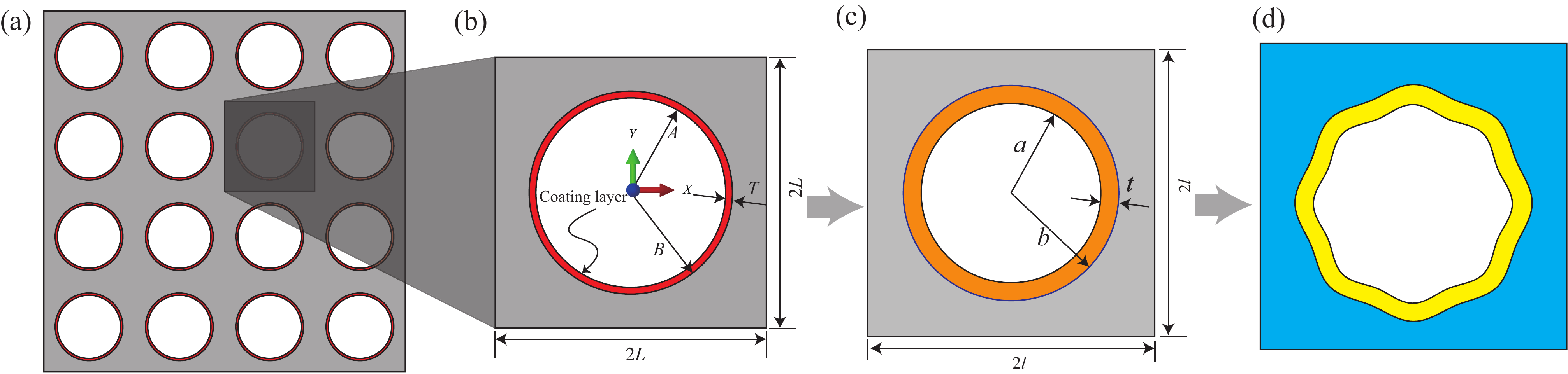}
\caption{(Color online) (a) A $4\times4$ structure in the periodically porous elastomer with stiff surface coating on each hole. (b) The chosen representative volume element with periodic boundary conditions. (c) Axisymmetric plane-strain deformation incurred by thermal loading. (d) Surface wrinkling occurs as the temperature exceeds a critical value.}\label{fig1}
\end{figure*}

In this work, a periodically porous elastomer with square arrays of circular holes is considered. In particular, each hole contains a surface coating, and the geometry of the structure is shown in Figure \ref{fig1}(a). Since the size of the original structure is infinite, we select a representative volume element (RVE) or a unit cell to carry out deformation analysis, and it is composed of a soft matrix (gray part) and an annulus layer (red part), as illustrated in Figure \ref{fig1}(b). In the initial state, the length of this specific unit cell is $2L$, and the inner and outer radii for the coating are denoted by $A$ and $B$, respectively. In addition, we use $T$ to represent the initial thickness of the coating layer. It is worth mentioning that a complete representative cell in most existing studies for porous elastomers is usually comprised of 2$\times$2 cells, and a global buckling pattern can be induced by compression, deflation, and so on \citep{cj2018,lwc2019,nly2020}. We define a mode that the boundary of the unit cell buckles as global buckling. In this study, surface instability in a unit cell may be incurred by contactless loading approaches, e.g., swelling, heating, magnetic field, light, etc. Here the so-called surface instability is similar to that in growth-induced surface wrinkling in tubular tissues and the instability is mainly concentrated on the annulus layer and the interface between the layer and the matrix. As a paradigmatic scenario,  we apply a thermal load to surface coating and assume that the metamaterial structure undergoes a plane-strain deformation. Furthermore, it is assumed that all interfaces between the coating layer and the matrix remain perfectly bonded in the deformation process. In this way, an axisymmetric deformation, shown in Figure \ref{fig1}(c), is first induced, and the critical geometrical parameters become $a$, $b$, $t$, and $l$, respectively. Finally, we anticipate that a surface wrinkling can emerge as the thermal strain passes a critical value (see Figure \ref{fig1}(d)). 

For convenience, we employ a two-dimensional rectangular coordinate system $\{O, X, Y\}$, and the origin is located at the center of the unit cell. In doing so, the boundaries of the unit cell in Figure \ref{fig1}(b) are given by $X=\pm L$ and $Y=\pm L$. Furthermore, it is expected that the selected unit cell can be used to analyze the deformation and instability of the whole structure when the following periodic boundary conditions are imposed:
\begin{equation}
\begin{split}
&U(L,Y)-U(-L,Y)=2L\varepsilon_{X},~~V(L,Y)-V(-L,Y)=0,\\
&U(X,L)-U(X,-L)=0,~~V(X,L)-V(X,-L)=2L\varepsilon_{Y},
\end{split}\label{eq2_1}
\end{equation}
where $U$ and $V$ signify the displacements in the $X$- and $Y$-directions, respectively, while $\varepsilon_{X}$ and $\varepsilon_{Y}$ the corresponding macroscopic mean strains. In this problem, we make use of a parameter $\varepsilon=(l-L)/L$ to clearly depict how large the unit cell expands.Moreover, thermal expansion of the coating is isotropic (possesses cubic symmetry). This ensures that the unit cell will constantly remain the square shape at any deformed configuration, thereby we acquire $\varepsilon_{X}=\varepsilon_{Y}=\varepsilon$.

It is assumed that both the matrix and the coating layer are composed of the neo-Hookean material, and the strain-energy function is given by
\begin{align}
W=\dfrac{\mu}{2}\left(\dfrac{\operatorname{tr}\mathbf{B}}{\sqrt[3]{\operatorname{det}\mathbf{B}}}-3\right)+\dfrac{\kappa}{2}\left(\sqrt{\operatorname{det}\mathbf{B}}-1\right)^2,\label{eq2_2}
\end{align} 
where $\mu$ and $\kappa$ stand for the shear modulus and bulk modulus, respectively, and $\mathbf{B}$ represents the left Cauchy-Green strain tensor. The Poisson's ratio $\nu$ is related to the shear and bulk moduli by 
\begin{align}
\nu=\dfrac{3\kappa-2\mu}{2(3\kappa+\mu)}.\label{eq2_3}
\end{align} 
It should be pointed out that an ideal incompressible material results in $\operatorname{det}\mathbf{B}=1$ so that the last term in (\ref{eq2_2}) vanishes. In this paper, we presume that the elastomer is comprised of rubber-like material which is almost incompressible. Since the designed metamaterial structure contains two materials, we shall put a ``~$\hat{~}$~'' on a notation if it signifies a physical quantity for the surface coating. For example, we let $\hat{\mu}$ denote the shear modulus for the coating. Yet if a quantity takes the same value in both the coating and matrix, we will omit the hat notation when it is evaluated in the coating area for the sake of brevity. 

In particular, the thin coating is stiffer compared to the matrix, indicating that the unit cell is analogue to a curved stiff film bonded to a compliant substrate. Based on previous studies on pattern formation and evolution in film-substrate structures, the ratio of the shear modulus of the coating layer to that of the matrix, which is denoted by $\xi$, as well as the initial thickness of the surface coating $T$ can both dramatically affect the buckling pattern \citep{ld2014,cf2019}. In detail, surface wrinkles can be triggered at a critical load only if the surface coating is slightly stiffer than the substrate. Otherwise, creasing profile with self-contact may occur instead \citep{ch2012b,cai2012}. For instance, a necessary condition for the appearance of surface wrinkling in a planar film coated to a half-space is $\xi>1.74$ \cite{cf1999,hutchinson2013}, while in the curved counterparts, this critical value of $\xi$ is a function of the thickness of the film \citep{jlc2019}. It is further deemed that different buckling patterns and buckling amplitudes may alter the band structures of the proposed metamaterial. Therefore, the current study aims at understanding a possibly novel buckling pattern as well as the wave propagation in porous elastomers via finite element (FE) simulations. 
%%%%%%%%%%%%%%%%%%%%%%%%%%%%%%%%%%%%%%%%%%%%%%%%%%%%%%%%%%%%%%%%%%%%%
\section{Buckling pattern induced by thermal loading}
In this section, we shall perform an FE analysis in the commercial software Abaqus in connection with deformation and instability for the $1\times 1$ unit cell. To realize thermal expansion, we need to set the thermal expansion coefficient for the surface coating. In doing so, the stiff coating will swell as temperature increases, and the soft matrix will deform as well due to the continuity condition on the interface. We emphasize that sinusoidal wrinkles in cylindrical structures are ubiquitously observed in nature and biological structures, for instance, gastrointestinal tract \citep{wms2012,bec2015}. Therefore, growth- or swelling-induced pattern formations have been widely studied in the literature. Correspondingly, the influence of material and geometric parameters, such as the shear modulus ratio $\xi$, has been unraveled using multiple approaches. We then briefly summarize the main conclusions. In a bilayer tube where each layer has its own shear modulus subjected to growth or swelling effect, a stiffer inner layer will reduce the critical load as well as the wavenumber \citep{lcf2011,mg2011}. In particular, by defining a critical thickness $t_{cr}$ where surface wrinkling initiates, we acquire that $t_{cr}/T-1$ is of $O(\xi^{-2/3})$ \citep{jlc2018}. Meanwhile, a thinner inner layer will produce more wrinkles. Furthermore, the main purpose of this paper is to explore the possibility of generating a new buckling pattern in periodically porous elastomers with applications to bandgap manipulations. There are two main tasks in this section. The first is to induce surface wrinkles and the second is to validate the size of the unit cell. In view of these facts, we only present a representative example in the following analysis.

In our illustrative calculation, the geometric parameters of the $1\times 1$ unit cell are given by $L=10$ mm, $A=6$ mm, and $B=6.1$ mm. For the material parameters, it is known that natural rubber is nearly incompressible. As mentioned earlier, we adopt a rubber-like structure and then specify the Poisson's ratios for the coating and matrix by $\nu=0.4997$. Moreover, the shear modulus of the soft matrix is specified by $\mu=1.08$ MPa \citep{wsb2013}. The bulk modulus can be determined by (\ref{eq2_3}) and is given by $\kappa\approx2$ GPa. In order to generate surface wrinkles, we take $\xi=300$ as a prototypical value. Under this situation, we obtain $\hat{\mu}=300\mu$ and $\hat{\kappa}=300\kappa$. 

\begin{figure}[!h]
\centering\includegraphics[width=9cm]{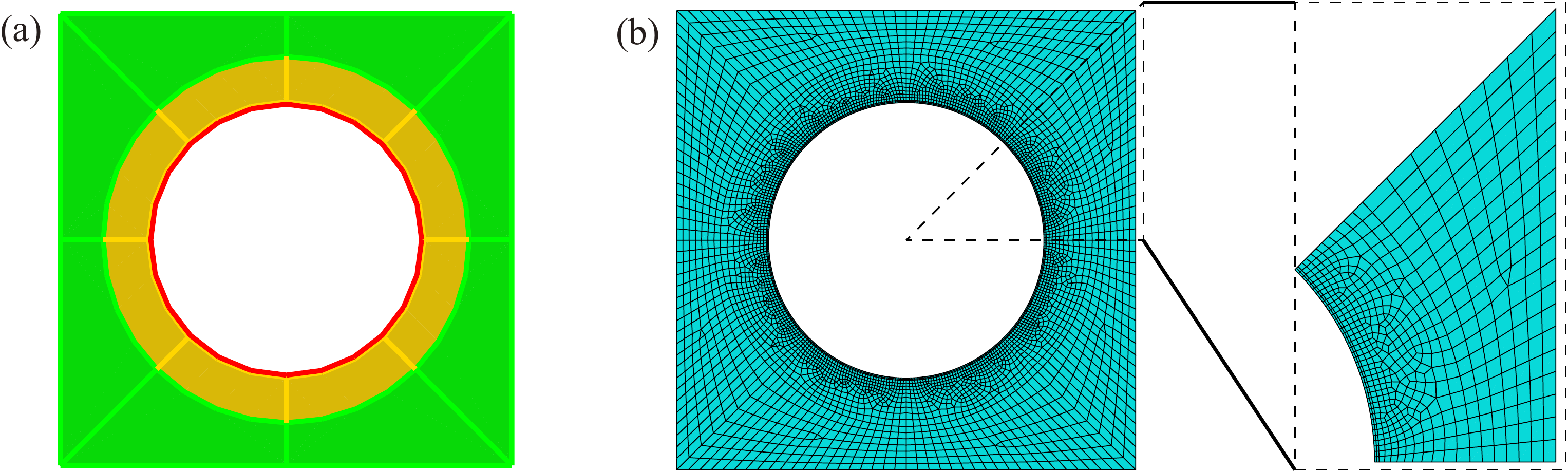}
\caption{(Color online) The non-uniform mesh strategy for the $1\times1$ unit. (a) The structure is partitioned into three regions marked by different colors. (b) The meshed unit cell with a blow-up.}\label{fig2}
\end{figure}

Subsequently, we perform a nonlinear analysis for the $1\times 1$ unit cell with the use of the built-in module ``Static, General'' in Abaqus. The periodic boundary conditions in (\ref{eq2_1}) can be implemented by secondary development of Abaqus using Python codes. On the one hand, to ensure the success of the periodic boundary conditions, the grid numbers of the four boundaries (two vertical lines and two horizontal lines) must be identical. On the other hand, it is anticipated that the deformation is practically homogeneous at the position far from the interface. Therefore, a non-uniform mesh strategy where the density of the grid decays from the innermost part is adopted in our FE simulation, which enables us to compute successfully and efficiently. We then shortly outline how we discretize the unit cell, and a graphical description is exhibited in Figure \ref{fig2}. It can be seen from Figure \ref{fig2}(a) that the $1\times1$ unit cell has been divided into three parts. The red area is the surface coating where the eight-node quadratic plane-strain elements with reduced integration (CPE8RH) are applied. The soft matrix contains two parts. The CPE8RH mesh type is used in the green region while a hybrid-mesh scenario is employed in the orange part (transition zone). In addition to the CPE8RH mesh type, the six-node quadratic plane-strain elements (CPE6H) are also adopted. In this way, we ensure that the FE model can remain a desired precision as well as an acceptable computation cost. We plot the meshed model in Figure \ref{fig2}(b). The number of CPE8RH mesh is 9046 and the counterpart of CPE6H is 104. 

In order to verify our Python scripts of periodic boundary conditions, we have reproduced the buckling pattern in \cite{wcs2014} by taking the same model, unit cell and parameters. As the compressive strain increases from 0 to 0.1, we obtain that the post-buckling results and the corresponding distributions of the normalized von Mises stress are the same as those shown in Figure 2(a) in \cite{wcs2014}.

\begin{figure*}[!htp]
\centering\includegraphics[width=15cm]{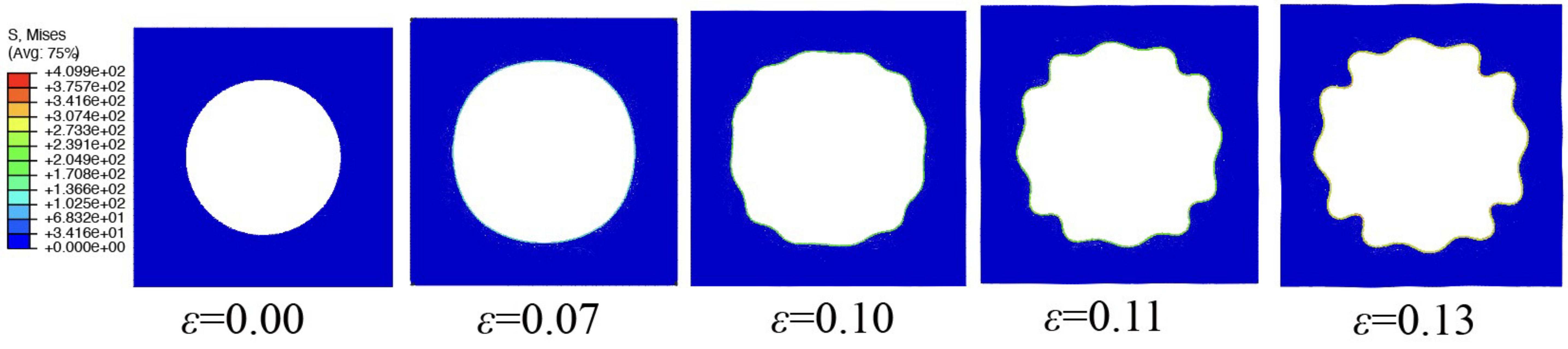}
\caption{(Color online) Deformation and surface wrinkling in the $1\times1$ unit cell at different values of the macroscopic mean strain $\varepsilon$.}\label{fig3}
\end{figure*}

As all preliminary settings mentioned above are well prepared, we carry out an FE simulation on heating-induced deformation and pattern formation in the $1\times1$ unit cell with periodic boundary conditions on all outside edges. As introduced earlier, we have taken $\varepsilon=(l-L)/L$ signifying the macroscopic mean strain as the bifurcation parameter since it is more convenient to depict how large the unit cell expands. Note that there exists a circular cavity, it is not necessary to introduce an additional geometrical or physical imperfection in the simulation as the discretization procedure is equivalent to approximating an ideal circle by a regular polygon and this can be viewed as a well-existed geometrical imperfection. Then we track the deformation process in Abaqus and display several snapshots in Figure \ref{fig3}. It is found that the surface coating deforms under temperature increase and further drives the elongation of the unit cell in both the vertical and horizontal directions. In particular, the soft matrix remains constantly the square geometry.  Meanwhile, the internal cavity keeps the circular shape until the macroscopic mean strain exceeds a critical value $\varepsilon_{cr}=0.0798$ where surface wrinkling initiates. There are in total 12 waves produced by thermal stress. Later, the amplitude of the sinusoidal wrinkles augments as the strain enlarges, and the four peaks near the diagonal lines quickly become higher, giving rise to a period-tripling mode. The deformed configurations of $\varepsilon=0.11$ and 0.13 in Figure \ref{fig3} depict the period-tripling pattern.

Next, we plan to verify that the $1\times1$ square unit is indeed the smallest representative volume element that can be used to model the deformation and instability of the whole infinite structure. For that purpose, we shall compare the deformations, the stress distributions, as well as the critical strains for the $1\times1$, $2\times2$, $4\times4$, and $8\times8$ structures. To this end, FE simulations are conducted for these structures in Abaqus as well. We establish the last three models based on the $1\times1$ element by the mirroring method in Abaqus. For instance, the $2\times2$ units contain in total four $1\times1$ elements, and the geometry of each element is identical to that in Figure \ref{fig1}(b). Moreover, the material parameters, the loading approach, as well as the mesh strategy, are the same as those for the $1\times1$ unit cell. Note that the grids in the $8\times8$ cells are relatively coarser compared to other cases in order to save computation cost. For the $2\times2$ and $4\times4$ units, the periodic boundary conditions are imposed on the four boundaries. Specifically, it is assumed that the $8\times8$ units with simply supported boundary conditions can be used to understand the deformation and instability of the periodically porous elastomer. Therefore, we directly fix the lower left corner and restrict the vertical movement of the lower right position and the horizontal displacement of the upper left location. In addition to these constraints, all other boundaries are free. Correspondingly, we are concerned with the units around the center where the boundary effect can be ignored.

\begin{figure}[!h]
\centering
\begin{minipage}[c]{.35\textwidth} 
\centering\includegraphics[scale=0.28]{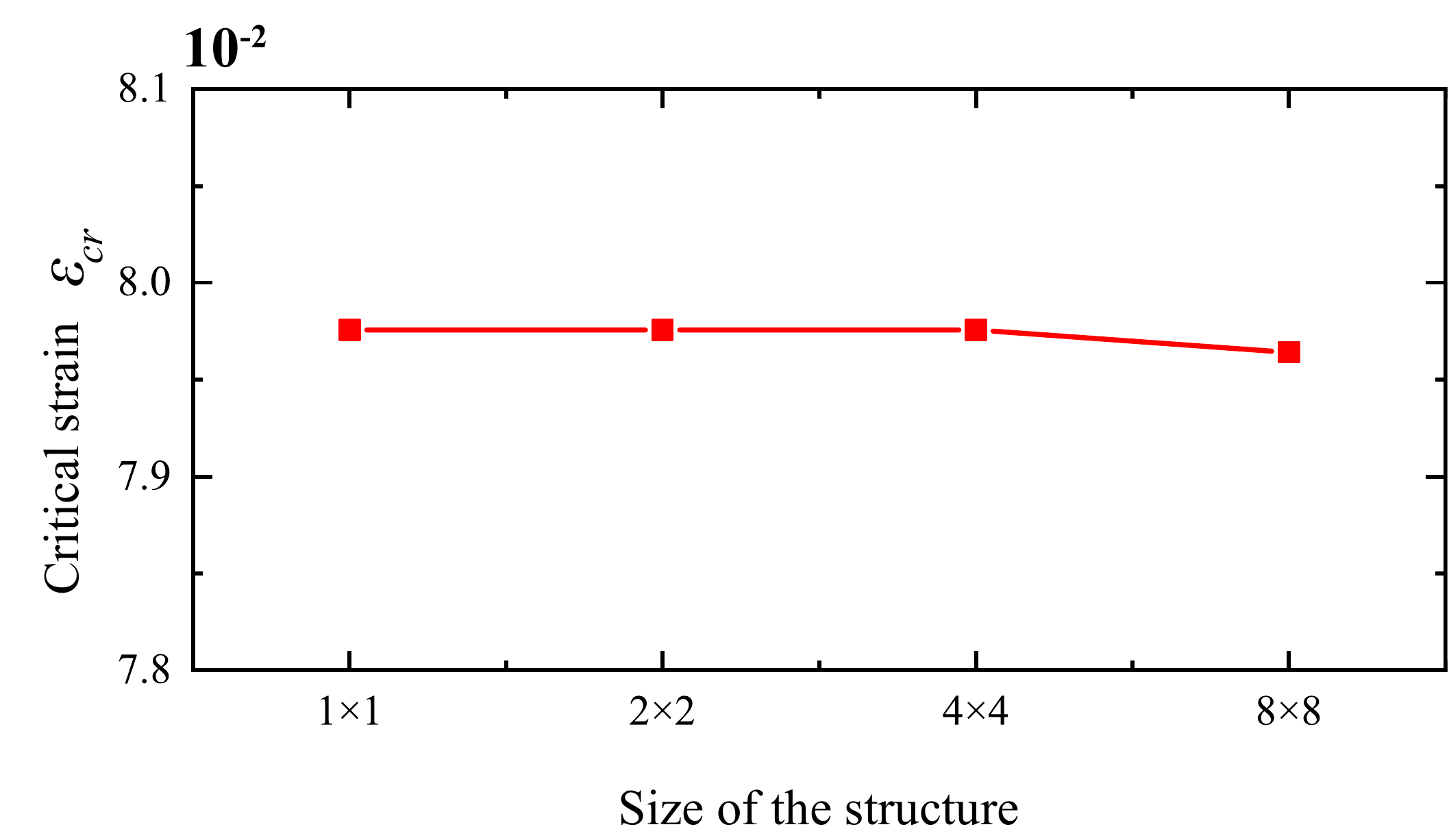}
\end{minipage}
\hfill 
\begin{minipage}[c]{.12\textwidth} 
\centering
\caption{(Color online) The dependence of the critical strain $\varepsilon_{cr}$ on the size of the unit cell.}\label{fig4}
\end{minipage}
\end{figure}

\begin{figure}[!h]
\centering\includegraphics[width=8cm]{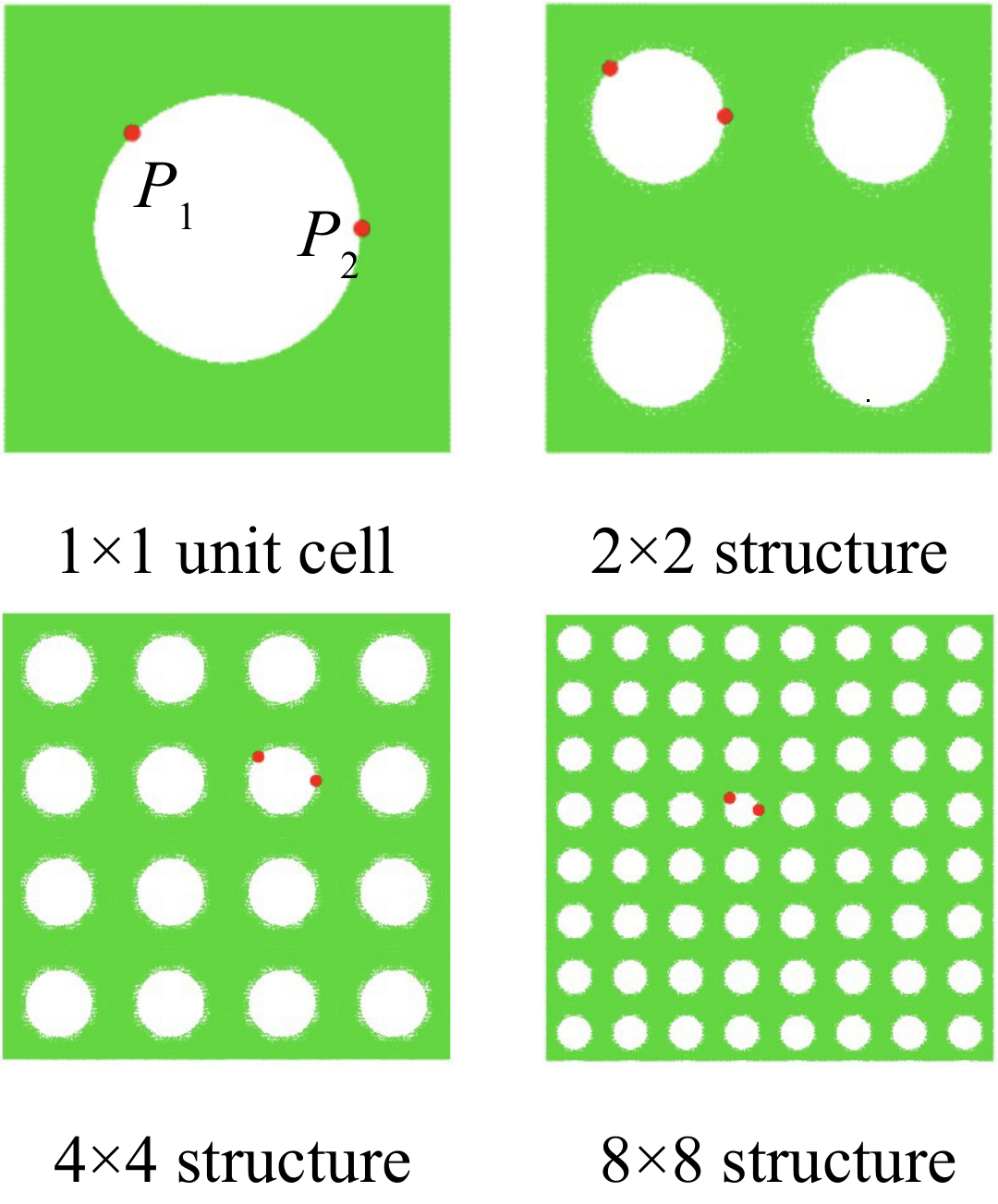}
\caption{(Color online) Positions of two arbitrarily chosen points $P_1$ and $P_2$.}\label{fig5}
\end{figure}

The critical strains for the selected four structures are illustrated in Figure \ref{fig4}. It turns out that the value of $\varepsilon_{cr}$ is nearly independent of the size of the structure. The critical strain for the $8\times8$ structure is slightly lower. This small deviation may be caused by the coarse mesh and the influence of the boundary effect since $8\times8$ structure with simply supported boundaries actually possesses a finite size. Furthermore, it is expected that the boundary effect decays rapidly such that the deformation and stress states in the internal part are no longer affected. We then randomly take two nodes $P_1$ and $P_2$ in the surface coating and their positions in different elements are shown in Figure \ref{fig5}. This selection is caused by the fact that the circular coating is 300 times stiffer than the matrix so that the stress is mainly concentrated on the surface. In detail, these two points locate on the same circle where the radius is given by $(A+B)/2$. In the $8\times8$ structure, the positions of  $P_1$ and $P_2$ lie in the center part in order to eliminate the boundary effect.

\begin{figure}[!h]
\centering
\subfigure[Stress of the $P_1$ point.]{\includegraphics[scale=0.2]{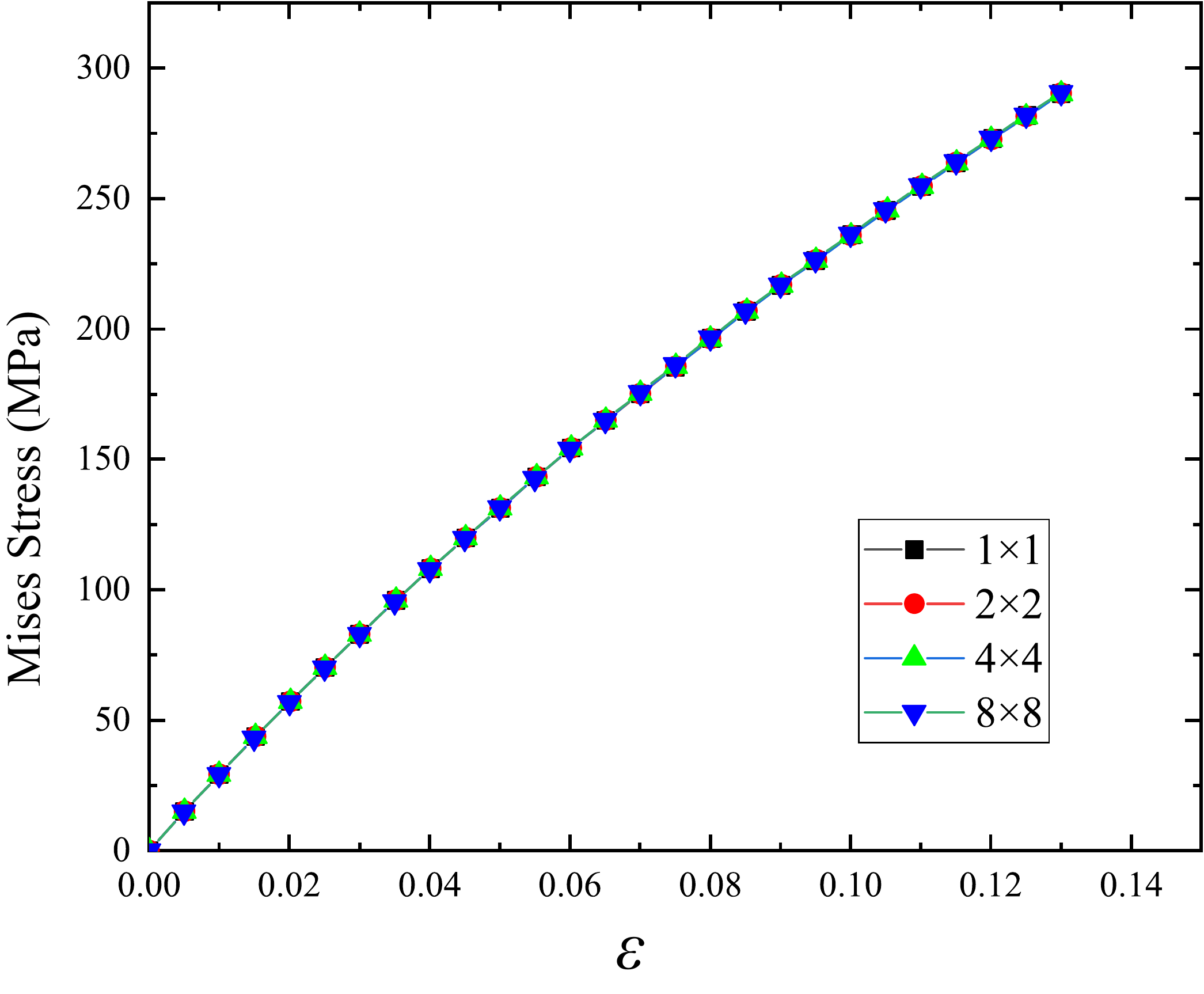}}
\subfigure[Stress of the $P_2$ point.]{\includegraphics[scale=0.2]{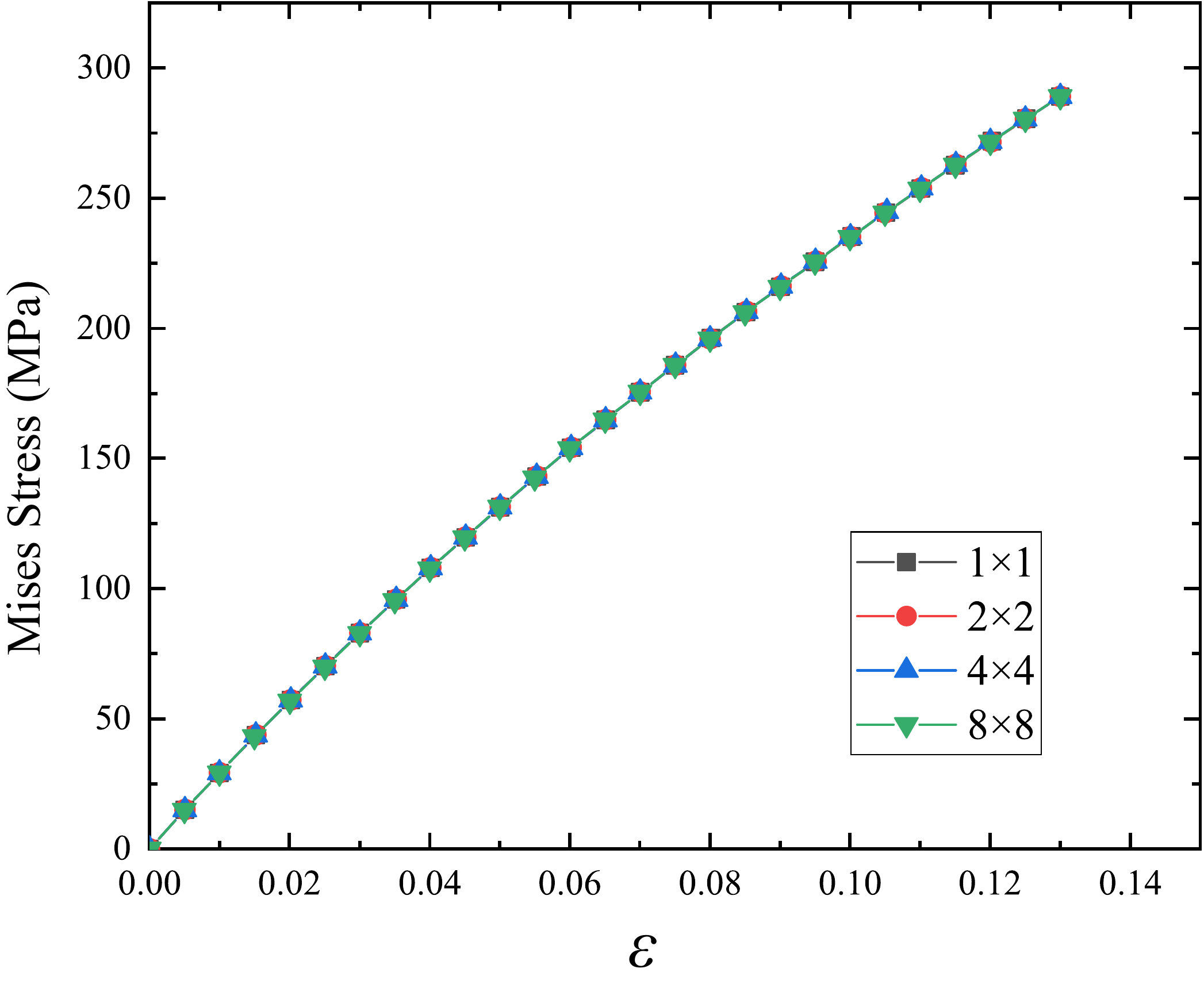}}
\caption{(Color online) The dependence of the von Mises stress on the macroscopic mean strain $\varepsilon$ for different sizes of the element.}\label{fig6}
\end{figure}

\begin{figure}[!h]
\centering\includegraphics[scale=0.38]{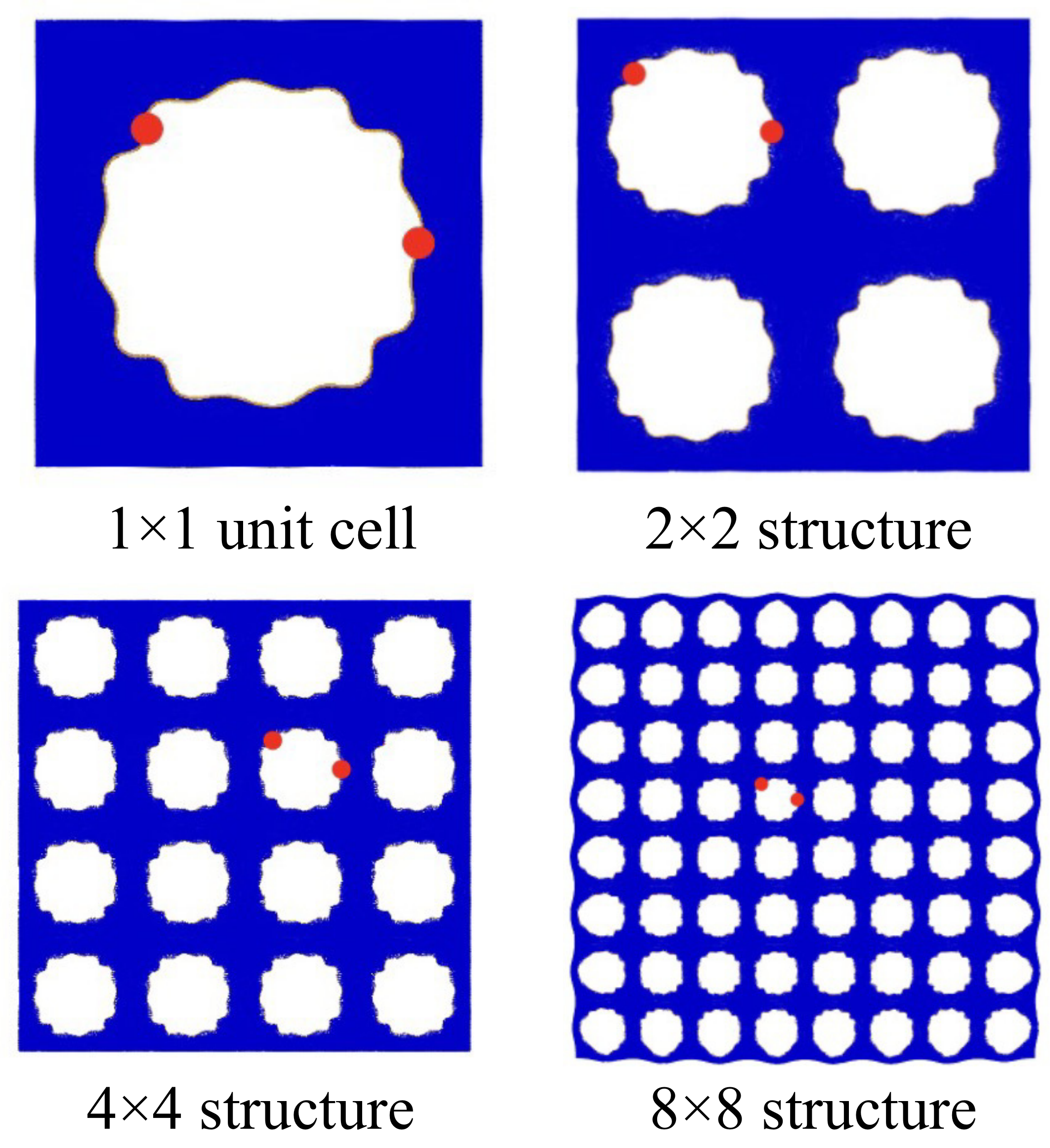}
\caption{(Color online) Post-buckling states for different cells when $\varepsilon=0.11$. The red dots highlight the positions of $P_1$ and $P_2$.}\label{fig7}
\end{figure}

Figure \ref{fig6} plots the von Mises stresses at $P_1$ and $P_2$ as functions of the macroscopic mean strain $\varepsilon$ for these four structures. In particular, the data exported from Abaqus are presented by distinct points, and the four curves agree extremely well. Finally, we show the deformed configurations for the chosen elements in Figure \ref{fig7} when $\varepsilon=0.11$. The locations of $P_1$ and $P_2$ are marked by red points. It is observed that the buckling pattern of the $1\times1$ unit cell is consistent with that appeared in any of the other three structures. In view of the comparisons illustrated in Figures \ref{fig4}, \ref{fig6}, and \ref{fig7}, we can conclude that the $1\times1$ unit cell does correspond to the RVE that can be used to characterize the deformation and instability for the metamaterial structure designed in this paper.

Currently, we have preliminarily unraveled a novel buckling pattern in a periodically porous elastomer where each cavity is covered by a stiff coating. The deformation can be generated by some non-contact loads and we take heating as a typical case study. It turns out that surface wrinkling may appear at a critical load and the RVE is composed of a $1\times1$ unit cell with the periodic boundary conditions. From the results for hard materials \citep{lnl2020}, a wrinkled pattern in a hole may amplify the effect of Bragg scattering, resulting in bandgaps of high frequencies. On the other hand, it is well known that pre-stress and buckling instability can dramatically affect wave propagation in metamaterials composed of soft elastomers. Thus, we intend to study the dynamical response of the proposed metamaterial structure by considering the new buckling pattern.

\section{Wave propagation}
In this section, we shall investigate the possibility of using buckling pattern of the new metamaterial as a medium to regulate wave propagation. In general, solving the dispersion relation for a deformed solid will lead to an eigenvalue problem of the dynamical equilibrium equation. We then simply summarize the incremental theory of a pre-stressed body by taking a problem subjected to a mechanical load as a paradigm. Suppose that the position vector of a material point in a continuum body is denoted by $\widetilde{\bm X}$ in its reference configuration $\mathcal{B}_0$. This vector becomes $\bm x$ in the current configuration $\mathcal{B}_t$ through the motion $\bm x(\widetilde{\bm X},t)$  where $t$ stands for the time variable. The deformation gradient from $\mathcal{B}_0\longrightarrow \mathcal{B}_t$ is given by
\begin{align}
\mathbf{F}=\operatorname{Grad} \bm x=\dfrac{\partial \bm x}{\partial\widetilde{\bm X}}.
\end{align}

Denoting the nominal stress by $\mathbf{S}$ and the density in the reference state by $\rho_0$, we can write the equation of motion in the absence of body force as follows
\begin{align}
\operatorname{Div} \mathbf{S}-\rho_0 \ddot{\bm x}=\bm 0,
\end{align}
where a dot on a quantity signifies the material time derivative. We mention that an operator with a capital letter will be evaluated in the reference configuration while the counterpart in lowercase will be calculated in the current configuration. For a compressible (or nearly incompressible) hyperelastic material, the nominal stress $\mathbf{S}$ can be determined by
\begin{align}
\mathbf{S}=\dfrac{\partial W}{\partial \mathbf{F}},
\end{align}
where the strain-energy function $W$ depends on $\mathbf{F}$.

Assuming that an infinitesimal but time-dependent displacement field $\bm u$ is superimposed on $\mathcal{B}_t$, we can express the dynamical incremental equation relative to the incremental stress $\bm \chi$ as follows
\begin{align}
\operatorname{div} \bm\chi^\textrm{T}-\rho \ddot{\bm x}=\bm 0,\label{eq4_4}
\end{align}
where $\rho=J^{-1}\rho_0$ signifies the current density and $J=\operatorname{det}\mathbf{F}$ denotes the volume change. In particular, the incremental stress $\bm\chi$ can be expanded in $\mathbf{F}$ and the resulting linearized expression reads
\begin{align}
\bm\chi=\mathcal{A}:\bm\eta, \textrm{~~~or~in~component~form~~} \chi_{ij}=\mathcal{A}_{jilk}\eta_{kl},
\end{align}
where $\bm\eta=\operatorname{grad} \bm u$  and $\mathcal{A}_{jilk}$ are the instantaneous moduli defined by \citep{ogden1984,ylf2020}
\begin{align}
\mathcal{A}_{jilk}=J^{-1}F_{jN}F_{lP}\dfrac{\partial^2 W}{\partial F_{iN}F_{kP}}.
\end{align}
Note that the Einstein summation convention has been used in the above equations.

We seek a solution of (\ref{eq4_4}) in the following form
\begin{align}
\bm u(\bm x,t)=\bm w(\bm x)\textrm{e}^{-\textrm{i}\omega t},\label{eq4_7}
\end{align}
where $\bm w(\bm x)$ is an unknown function and $\omega$ expresses the angular frequency of an elastic wave. Indeed, this solution corresponds to a harmonic wave. In a similar manner, the incremental stress $\bm \chi$ can be rewritten as
\begin{align}
\bm\chi(\bm x,t)=\bm\Sigma(\bm x)\textrm{e}^{-\textrm{i}\omega t},\label{eq4_8}
\end{align}
with $\bm\Sigma(\bm x)= \mathcal{A}:\operatorname{grad}\bm w(\bm x)$. Substituting equations (\ref{eq4_7}) and (\ref{eq4_8}) into (\ref{eq4_4}) yields 
\begin{align}
\operatorname{div} \bm\Sigma^\textrm{T}+\rho\omega^2\bm w=\bm 0.
\end{align}
It can be seen that the problem is now transferred into a differential equation with respect to $\bm w$.

For a two-dimensional periodic structure where elastic waves propagate, it is convenient to focus on a unit cell which can be defined by a set of base vectors $\bm g_i$ $(i=1,2)$. In doing so, any particle $\bm$ in the periodic structure can be related to the unit cell through 
\begin{align}
\bm x=n_i \bm g_i+\bm x_0,
\end{align}
where $n_1$ and $n_2$ are arbitrary integers, and $\bm x_0$ represents the corresponding position vector within the unit cell. In addition, we define the reciprocal space (or wave vector space) of the  unit cell by two vectors $\bm g^j$ $(j=1,2)$ satisfying the following relation
\begin{align}
\bm g_i \cdot\bm g^j=2\pi\delta_{ij},
\end{align}
with $\delta$ being the Kronecker delta. It then follows that 
\begin{align}
\bm g^1=2\pi\dfrac{\bm g_2\times(\bm g_1\times \bm g_2)}{(\bm g_1\times \bm g_2)\cdot(\bm g_1\times \bm g_2)},~\bm g^2=2\pi\dfrac{(\bm g_1\times \bm g_2)\times\bm g_1}{(\bm g_1\times \bm g_2)\cdot(\bm g_1\times \bm g_2)}.
\end{align}

According to the Floquet-Bloch theory, the following relations can be obtained
\begin{align}
\begin{split}
\bm w(\bm x_0+\bm T)=\bm w(\bm x_0)\textrm{e}^{\textrm{i} \bm k \bm T},~~\bm \Sigma(\bm x_0+\bm T)=\bm \Sigma(\bm x_0)\textrm{e}^{\textrm{i}\bm k\bm T}
\end{split}\label{eq4_13}
\end{align}
where $\bm T=n_i \bm g_i$ signifies the periodicity of the structure and $\bm k$ denotes the wave vector in the reciprocal space. In particular, the reciprocal space also admits a periodicity and the wave vector $\bm k$ can be written as
\begin{align}
\bm k=m_j  \bm g^j+\bm k_0,
\end{align}
where $m_1$ and $m_2$ are also integers, and $\bm k_0$ is the wave vector in the unit cell. Accordingly, we acquire 
\begin{align}
\bm w(\bm x_0)\textrm{e}^{\textrm{i}\bm k \bm T}=\bm w(\bm x_0)\textrm{e}^{\textrm{i} \bm k_0 \bm T}.
\end{align}
 
Then we can solve the incremental equation for the unit cell associated with the Bloch boundary conditions to determine the dispersion relation $\omega(\bm k_0)$. Due to the periodicity, it is only necessary to compute the wave vectors on the boundaries of the irreducible Brillouin zone for a bandgap analysis. Specifically, in this problem, the base vectors for the unit cell take the specific forms $\bm g_i=2L\bm e_i$ $(i=1,2)$ with $\{\bm e_1,\bm e_2\}$ being the orthonormal basis for the two-dimensional Cartesian coordinates. In doing so, the reciprocal base vectors are given by $\bm g^j=\pi\bm e_j/L$ $(j=1,2)$. Bearing in mind that the unit cell preserves the square geometry in a buckled configuration, we may employ a triangle in the first quadrant as the irreducible Brillouin zone, which is shown in Figure \ref{fig8}. In fact, we find that dispersion curves for all eight triangles in Figure \ref{fig8} for a specific buckled state are identical, which verifies the selection of the irreducible Brillouin zone. We have marked three points by $G$, $E$, and $M$, respectively, and the boundaries of the irreducible Brillouin zone are obtained by joining them anti-clockwise. Then we shall calculate the wave vectors in the three lines $GE$, $EM$, and $MG$. It is pointed out that the Bloch conditions need to be implemented in Abaqus via Python coding to carry out wave analysis.

\begin{figure}[!h]
\centering
\begin{minipage}[c]{.3\textwidth}
\centering
\includegraphics[scale=0.3]{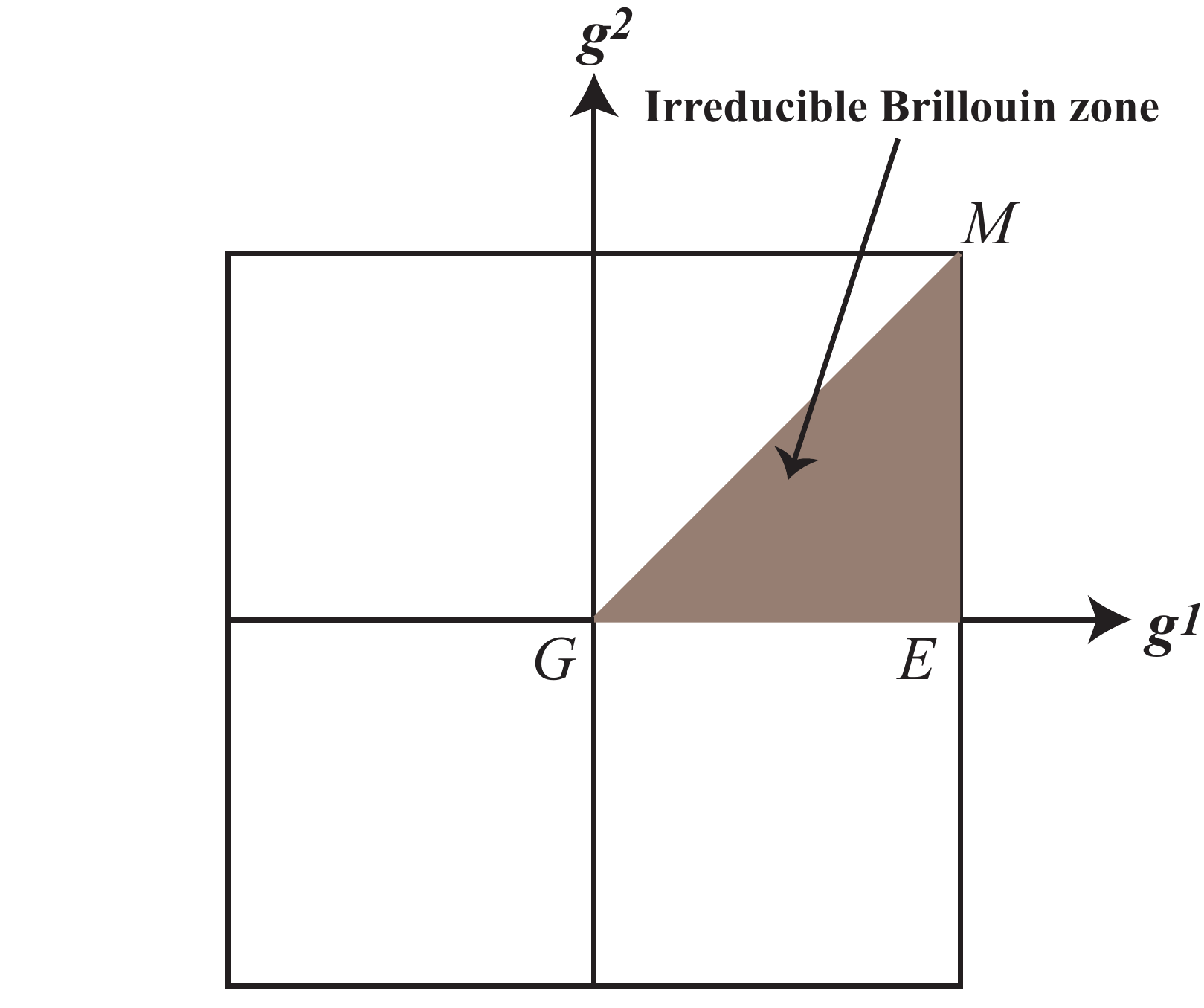}
\label{fig8}
\end{minipage}
\begin{minipage}[c]{.15\textwidth}
\centering
\caption{(Color online) The triangle $GEM$ signifies the irreducible Brillouin zone of a reciprocal lattice.}\label{fig8}
\end{minipage}
\end{figure}

Because Abaqus can not deal with complex-valued displacements which are required in the Bloch wave calculation, we shall split any  field $\phi$ (displacement or stress) into real and imaginary parts:
\begin{align}
\phi(\bm x)=\phi^{Re}(\bm x)+\textrm{i} \phi^{Im}(\bm x),
\end{align}
where the superscripts $Re$ and $Im$ imply the real and imaginary parts, respectively. By doing so, the incremental equation (\ref{eq4_4}) can also be divided into two sets of uncoupled equations for the real and imaginary parts, which are not shown here for brevity. On the boundaries of the unit cell, the displacement and traction are required to be continuous. Note that any traction on a prescribed surface with normal vector $\bm n$ can be identified as $\bm\chi \bm n=(\bm\Sigma(\bm x)\textrm{e}^{-\textrm{i}\omega t})\bm n=\bm q(\bm x_0)\textrm{e}^{-\textrm{i}\omega t}$. Utilizing equation (\ref{eq4_13}), the continuity conditions furnish \citep{ag1997}
\begin{align}
\begin{split}
&\bm w^{Re}(\bm x_b)=\bm w^{Re}(\bm x_b+\bm T)\operatorname{cos}(\bm k_0\bm T)+\bm w^{Im}(\bm x_b+\bm T)\operatorname{sin}(\bm k_0\bm T),\\
&\bm w^{Im}(\bm x_b)=\bm w^{Im}(\bm x_b+\bm T)\operatorname{cos}(\bm k_0\bm T)-\bm w^{Re}(\bm x_b+\bm T)\operatorname{sin}(\bm k_0\bm T),\\
&\bm q^{Re}(\bm x_b+\bm T)=-\bm q^{Re}(\bm x_b)\operatorname{cos}(\bm k_0\bm T)+\bm q^{Im}(\bm x_b)\operatorname{sin}(\bm k_0\bm T),\\
&\bm q^{Im}(\bm x_b+\bm T)=-\bm q^{Re}(\bm x_b)\operatorname{sin}(\bm k_0\bm T)-\bm q^{Im}(\bm x_b)\operatorname{cos}(\bm k_0\bm T).
\end{split}\label{eq4_17}
\end{align}
We have used $\bm x_b$ to indicate a position vector at the boundary in equation (\ref{eq4_17}), whereby all the real and imaginary parts of nodes at the boundary can be connected. Finally, both real and imaginary parts will give rise to the same equation as follows
\begin{equation}
(\bm{\Phi}-\omega^2\bm{\Pi})\bm{v}=\bm{f}_n
\end{equation}  
where $\bm{\Phi}$ and $\bm{{\Pi}}$ are the stiffness and mass matrices, respectively, $\bm v$ corresponds to the generalized nodal displacement vectors and $\bm f_n$ the generalized nodal force vectors. When $\bm k_0$ is prescribed, we can acquire a list of the frequencies $\omega$ and further determine the band structures of the unit cell. 

\begin{figure}[!h]
\centering
\subfigure[The undeformed state when $\varepsilon=0$.]{\includegraphics[scale=0.2]{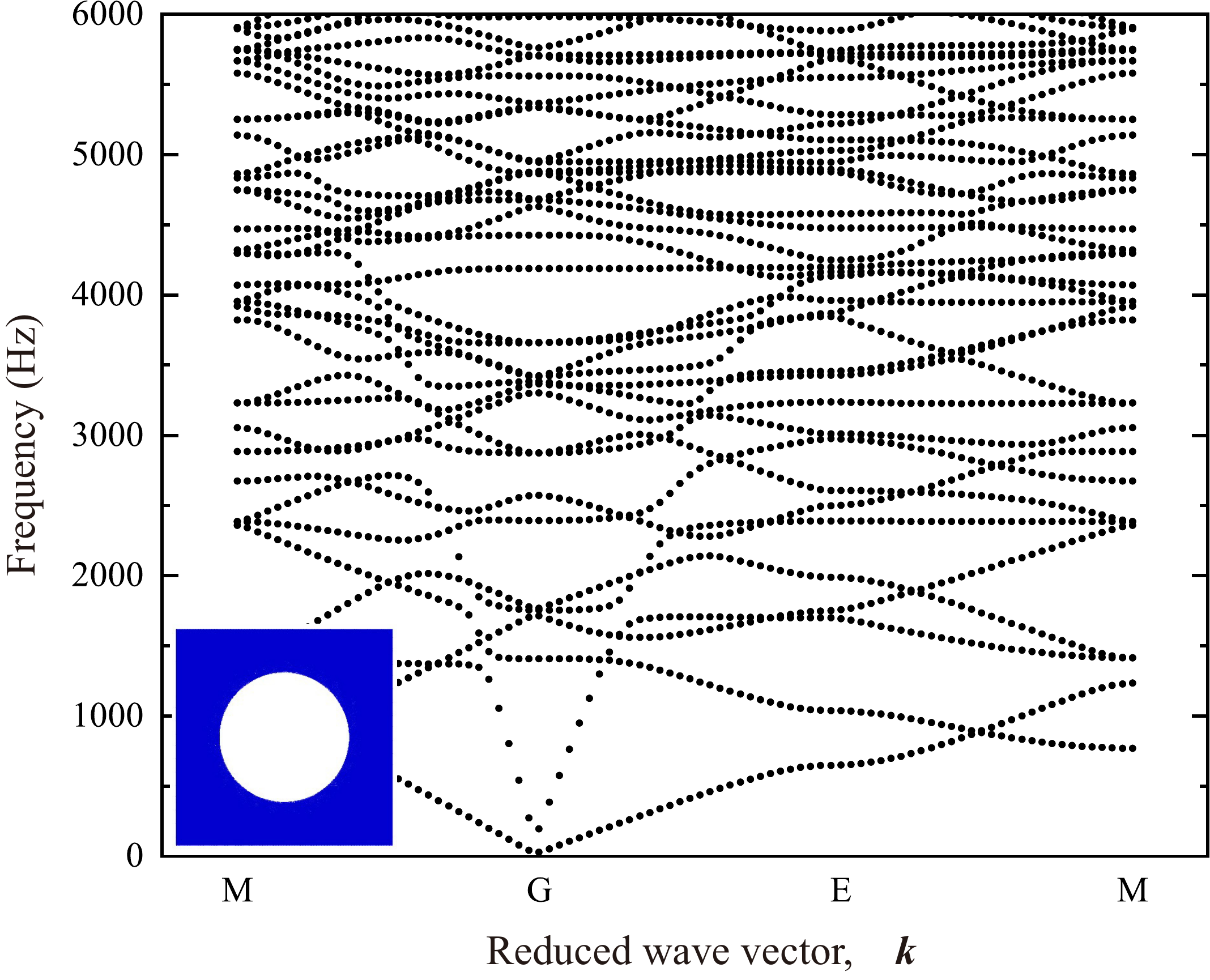}{\label{fig9a}}}
\subfigure[The buckled state when $\varepsilon=0.13$.]{\includegraphics[scale=0.2]{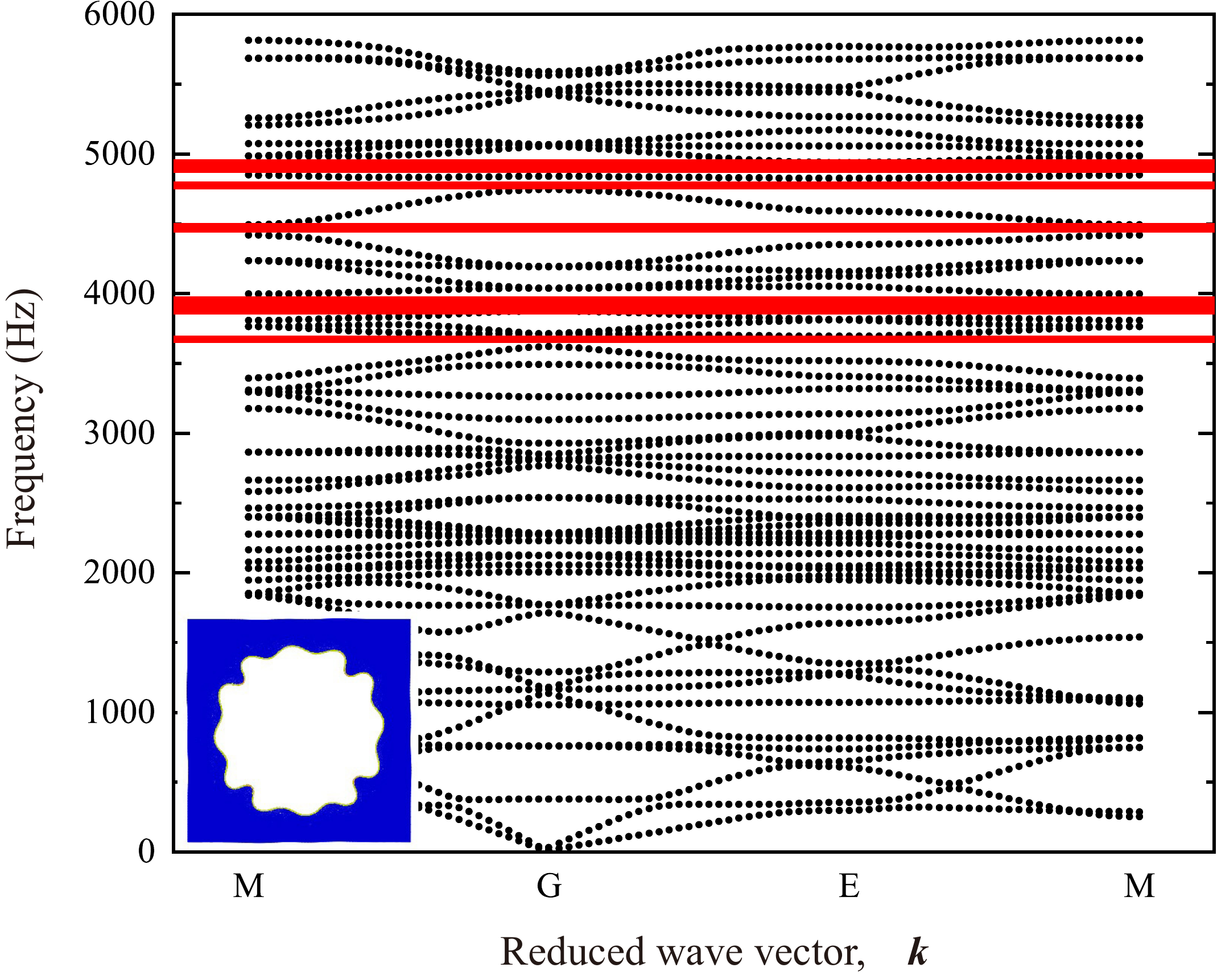}{\label{fig9b}}}
\caption{(Color online) Dispersion relationships for the unit cell when the Bloch boundary condition is employed. The undeformed and deformed states of the unit cell wherein an elastic wave propagates are also illustrated in each subfigure, and the area representing a bandgap has been highlighted by red color.}\label{fig9}
\end{figure}

Next, we employ the solution strategy implemented in Abaqus to study the dispersion relation of a pre-stressed state. Before proceeding further, we shall also examine the validity of our Python script for Bloch boundary conditions. To this end, the bandgaps in \citep{wsb2013} and \cite{glb2019b} are fully reproduced, although the later article used the software COMSOL. Without loss of generality, the initial geometrical and material parameters of the $1\times1$ unit cell are the same as those used in the calculation example in the previous section, namely, $L=10$ mm, $A=6$ mm, $B=6.1$ mm, $\nu=0.4997$, $\mu=1.08$ MPa, $\kappa=2$ GPa and $\xi=300$. In addition, the densities of both the surface coating and soft matrix in the undeformed configuration are specified by $\rho_0=1050$ $\textrm{kg}/\textrm{m}^3$. We refer to Figure \ref{fig3} for the deformation and buckling pattern associated with the macroscopic strain $\varepsilon$. In order to study how the buckling pattern influences the dispersion curves, we select two representative states as the base structures whereby elastic waves can propagate. Compared to the undeformed unit cell (also stress-free), the strain and stress states of a specific post-buckling configuration will be adopted as the initial state in order to obtain band structure of the deformed unit cell. The linear perturbation step is used to obtain the dispersion relationships and vibration modes under different wave speeds, and the results are exhibited in Figure \ref{fig9}. It is observed from Figure\ref{fig9a} that there is no bandgap in the stress-free state while five complete bandgaps appear in Figure \ref{fig9b} as the macroscopic strain attains $\varepsilon=0.13$. In addition, some nearly horizontal dispersion curves originate in certain frequency ranges (see also Figure \ref{fig12b}), which may lead to a trapped mode, possibly accompanied by local vibrations, as the group velocity becomes negligible. On the one hand, every bandgap occupies a narrow width. For instance, the widest one lies in the region  3875.3 Hz - 3996.3 Hz, and the gap width is nearly 121 Hz. On the other hand, all bandgaps are located on the high-frequency area. Indeed, this is in accordance with our expectations since the surface coating forms a wrinkled pattern which can induce Bragg scattering, and in principal Bragg scattering may generate bandgaps in high-frequency areas. In some practical applications, waves with low-frequency are expected to be prevented or attenuated. Meanwhile, it is known that local resonance is capable of producing complete low-frequency bandgaps \citep{lzm2000,zlh2012}. As a result, the add of an oscillator in each hole may enable us to pursue a wider bandgap at lower frequency ranges, and this will be explored in the next section.

%%%%%%%%%%%%%%%%%%%%%%%%%%%%%%%%%%%%
\section{An updated metamaterial structure with oscillators}
\begin{figure}[!h]
\centering\includegraphics[width=9cm]{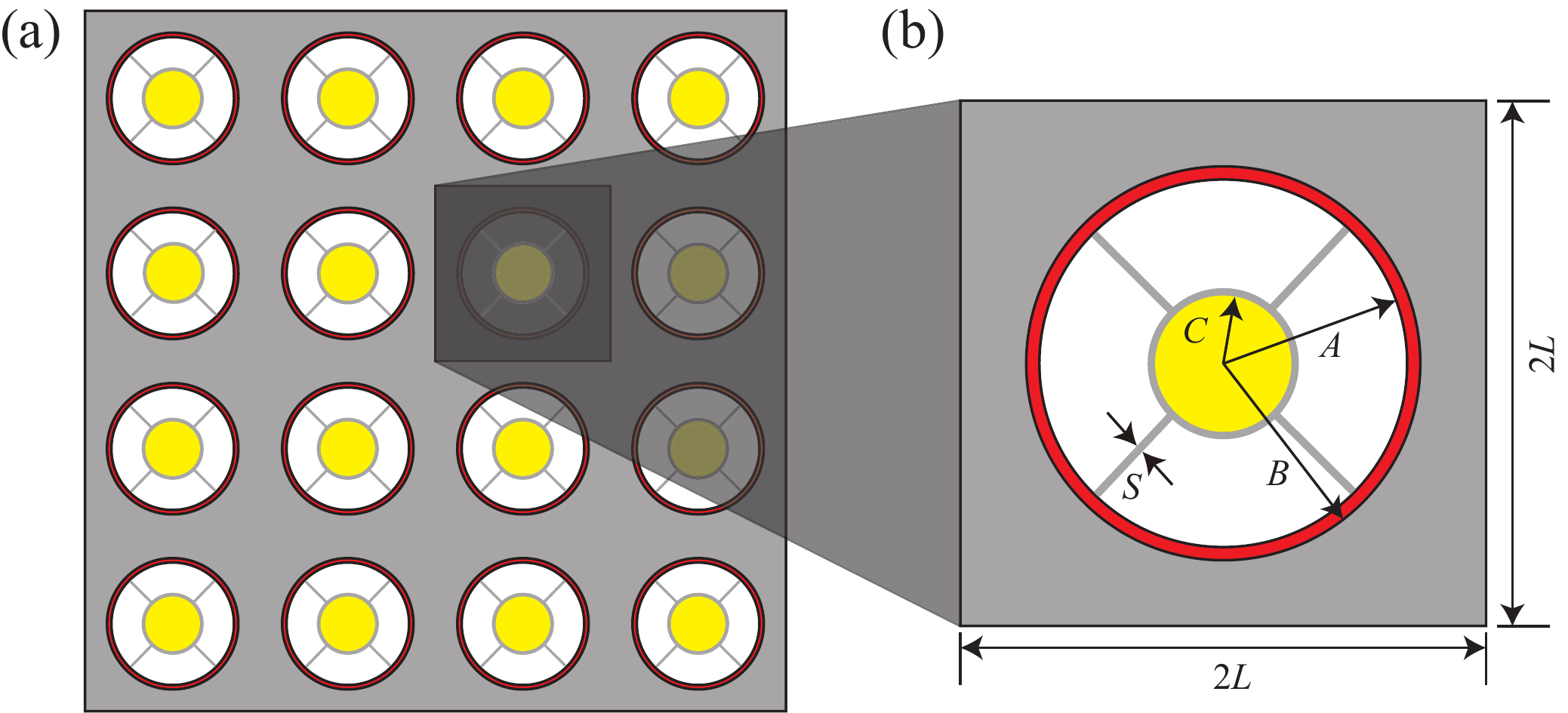}
\caption{(Color online) (a) A $4\times4$ structure in the periodically porous elastomer with stiff surface coating and an oscillator in each hole. (b) The $1\times1$ unit cell.}\label{fig10}
\end{figure}

Inspired by \cite{wcs2014}, we introduce a cylindrical core made of No.45 steel packed by diagonally arrayed soft beams in the unit cell of the original metamaterial structure, and the upgraded structure is displayed in Figure \ref{fig10}. Compared to the original structure, there are two additional geometrical parameters involved, i.e. the width of the soft beam $S$ and the radius of the metallic core $C$. It is noteworthy that the thickness of the annulus surrounding the steel circle is identical to the width of the beam, and the length of the soft beam connecting the surface coating and the oscillator is then given by $A-C-S$. Similarly, the unit cell is found to be a $1\times1$ unit by the same procedure adopted in the previous section. We have repeated the comparisons of the strains, stresses, and buckling patterns among the $1\times1$, $2\times2$, $4\times4$, and $8\times8$ units for different values of the macroscopic strain $\varepsilon$. In particular, different from the $8\times8$ structure which is simply supported, the periodic boundary conditions are applied to the other three structures. Notwithstanding, we omit more details for brevity and will directly employ the $1\times1$ unit as the unit cell in the following analysis. 

The deformation induced by heating can be traced via FE analysis in Abaqus (``Static, General'' module) by imposing the periodic boundary conditions (\ref{eq2_1}) on the unit cell. Likewise, the thermal load is only applied to the surface coating and the matrix is capable of finite deformation because of the fact that its link to the coating will be constantly retained. Furthermore, the metallic core is regarded as a rigid body. The soft beams, as well as the circular ring enclosing the oscillator, consist of the same material as that for the matrix. With an eye to comparing the deformation to the counterpart without oscillators shown in Figure \ref{fig3}, we let $L=10$ mm, $A=6$ mm, $B=6.1$ mm, $C=2$mm, $S=0.4$ mm, $\nu=0.4997$, $\mu=1.08$ MPa, $\kappa=2$ GPa, $\rho_0=1050$ $\textrm{kg}/\textrm{m}^3$ and $\xi=300$. As the metallic core is made of No. 45 steel, the Poisson's ratio $\nu_{st}$, the Young's modulus $E_{st}$, as well as the density $\rho_{st}$ are specified by $\nu_{st}=0.269$, $E_{st}=206$ GPa and $\rho_{st}=7890 $kg$/\textrm{m}^3$. The grid type and mesh generation can refer to the approach outlined in Section 3. The CPE8RH mesh type is applied for the soft beams and annulus while a combination of CPE8RH and CEP6H meshes is used for the metallic core.

\begin{figure*}[!h] 
\centering\includegraphics[width=17cm]{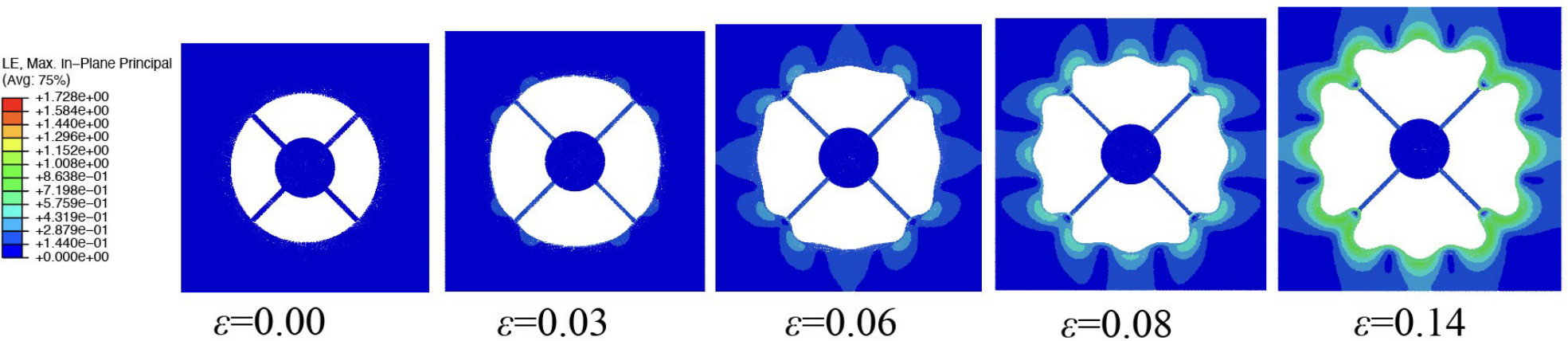}
\caption{(Color online) Pattern formation in the $1\times1$ unit cell at different values of the macroscopic mean strain $\varepsilon$ when a metallic core is embedded in the center hole.}\label{fig11}
\end{figure*} 

Figure \ref{fig11} exhibits several snapshots of the deformed configurations corresponding to distinct strain values of $\varepsilon$. The legend indicates the distribution of the logarithmic strain. Although the inner surface of the coating is no longer free, the constraints are symmetric with respect to the diagonal lines of the square cell. This difference causes an incompatible deformation in the hoop direction observed as $\varepsilon=0.03$. Because no external load is exerted on the soft beams, the free boundary of the surface coating expands more compared to the four joints where the dilation is constrained. It can be seen that the square matrix deforms driven by thermal expansion of the stiff coating, and the square geometry remains during the deformation. Initially, excepting the joints to the four stripes, the free surface keeps the circular shape as the deformation progresses until $\varepsilon$ exceeds practically a critical value 0.0346 where a wrinkled pattern is triggered. Furthermore, the surface wrinkles will deepen with increasing $\varepsilon$. Meanwhile, the soft beams are stretched and become thinner, and the diameter of the center hole enlarges continually. Note that the metallic core  is viewed as a rigid body and it keeps stationary due to the symmetry of the four beams. Parallel with the results in Figure \ref{fig3}, both the deformation and the buckling pattern are quite similar. However, the insertion of resonant cell may bring into novel wave properties, and this will be inspected in the subsequent analysis.

\begin{figure}[!h]
\centering
\subfigure[The undeformed state when $\varepsilon=0$.]{\includegraphics[scale=0.2]{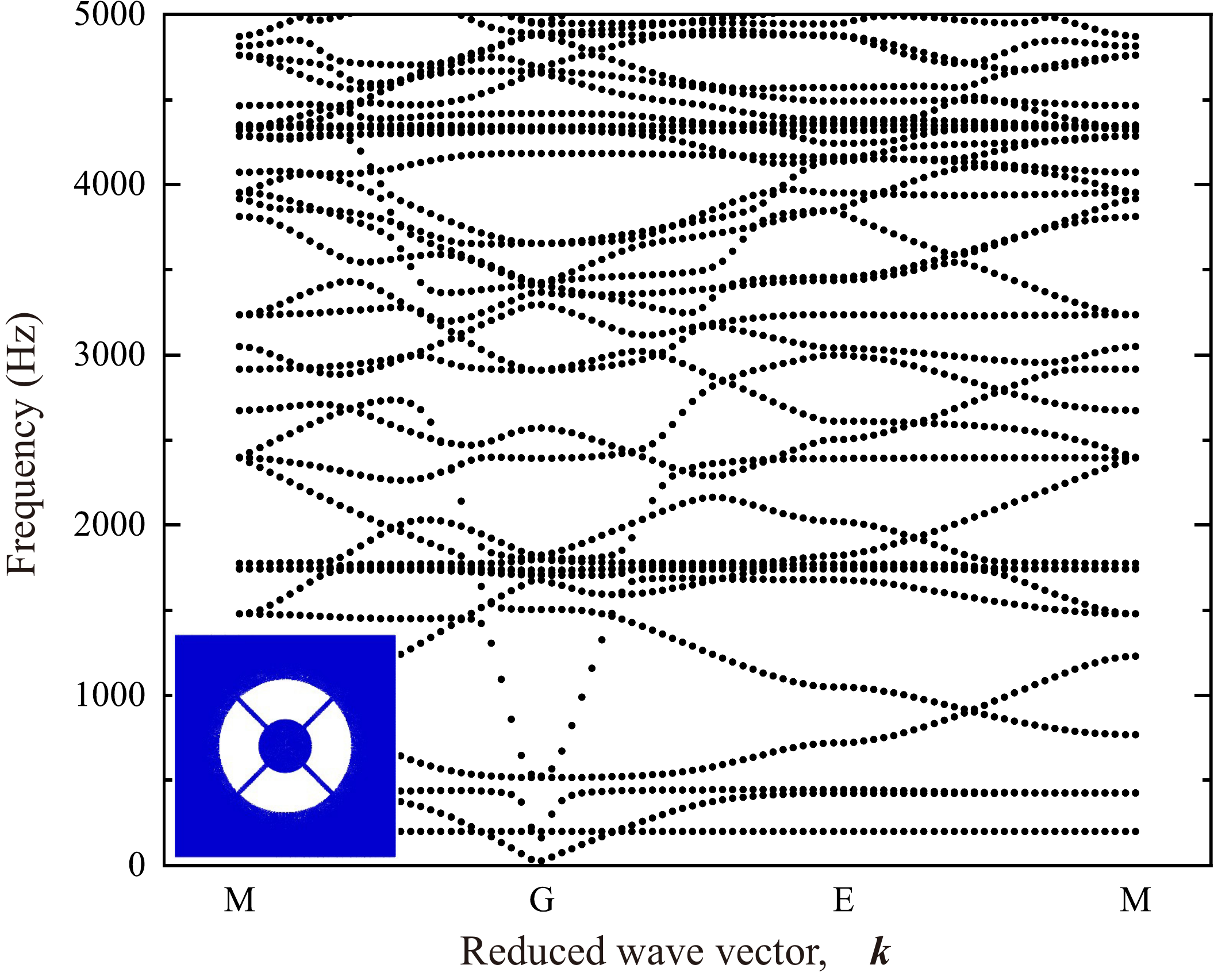}{\label{fig12a}}}
\subfigure[The buckled state when $\varepsilon=0.14$.]{\includegraphics[scale=0.2]{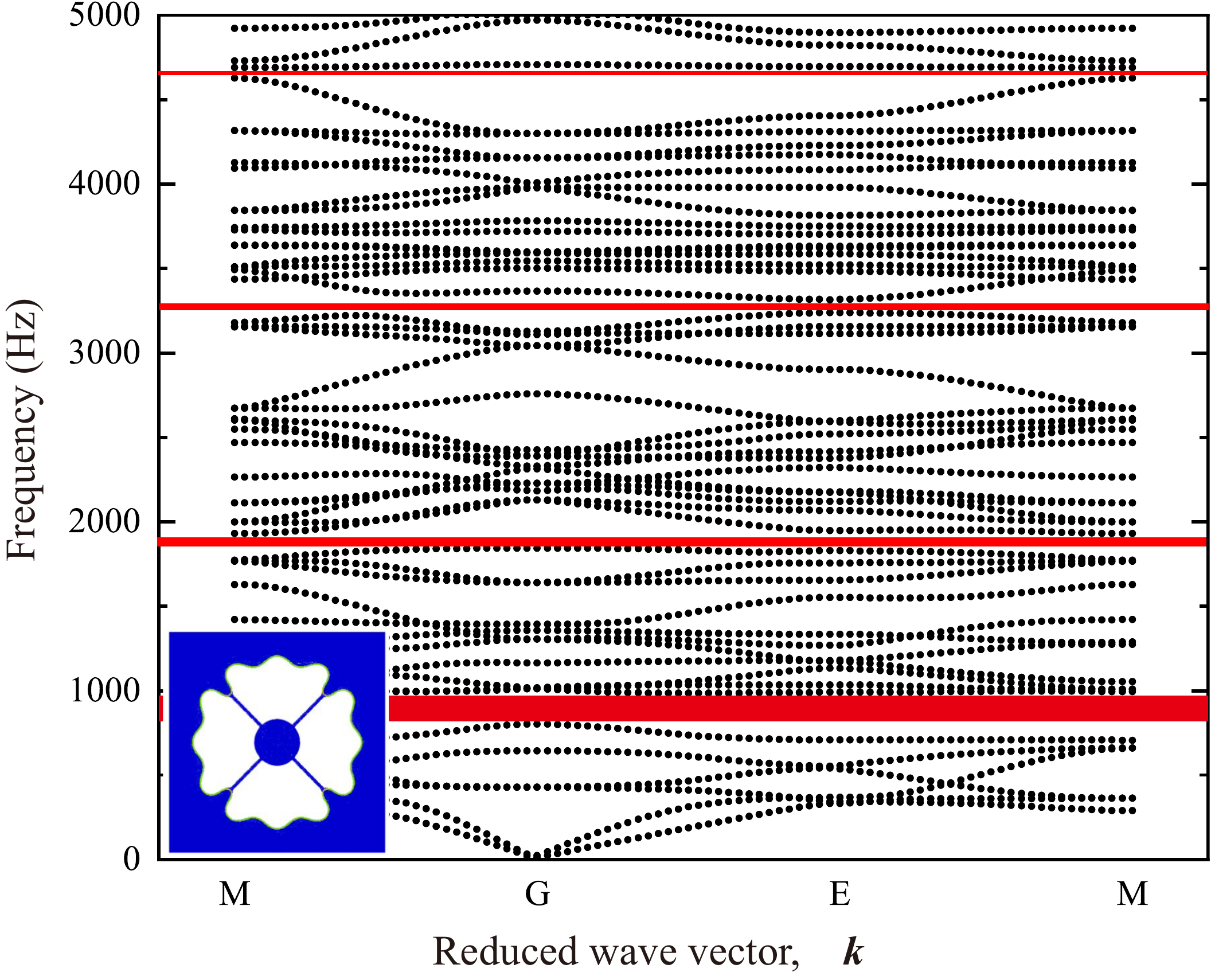}{\label{fig12b}}}
\caption{(Color online) Dispersion relationships for the unit cell when the Bloch boundary condition is employed. The undeformed and deformed states of the unit cell wherein an elastic wave propagates are also illustrated in each subfigure, and the bandgap has been highlighted by red color.}\label{fig12}
\end{figure}

Figure \ref{fig12} illustrates the dispersion relations for the initial configuration (stress-free state) and the buckled one when $\varepsilon=0.14$. All calculations are conducted in Abaqus. Analogue to the counterpart shown in Figure \ref{fig9a}, there is no bandgap in the undeformed state, as shown in Figure \ref{fig12a}. However, as the macroscopic strain $\varepsilon$ attains 0.14, several bandgaps are generated in Figure \ref{fig12b}. Although the bandgaps in the high frequency domain are narrow, a wider bandgap in the low-frequency range is found, which has not been observed in Figure \ref{fig9b}. This implies that the oscillator does alter the wave property of the structure and leads to the possibility of a bandgap of low frequency. In particular, the frequency range of this new bandgap is 801.95 Hz - 984.24 Hz (bandgap width 182.29 Hz). 

We further give another interpretation on the bandgap structure relative to the size of the unit cell. To this end, we compute the speed of longitudinal wave in the elastomer (soft matrix) by $v_l=\sqrt{\kappa/\rho_0}\approx1380.13~\textrm{m}/\textrm{s}$. Furthermore, we use $f_m$ to denote the average frequency of a bandgap. For the lower bandgap between 801.95 Hz - 984.24 Hz, the average value reads 893.095 Hz. The accompanied wavelength can then be determined by $\lambda_l=v_l/f_m\approx1545$ mm, which is nearly 80 times greater than the size of the unit cell ($2L=20$ mm). As pointed by Liu et al. \cite{lzm2000}, we refer to the widest bang gap in Figure \ref{fig12b} as a kind of low-frequency bandgap. In a word, it is possible to make use of such a structure with a relatively small characteristic scale (size of the unit cell) to control the sonic wave with large wavelength. 

\begin{figure*}[!h] 
\centering\includegraphics[width=17cm]{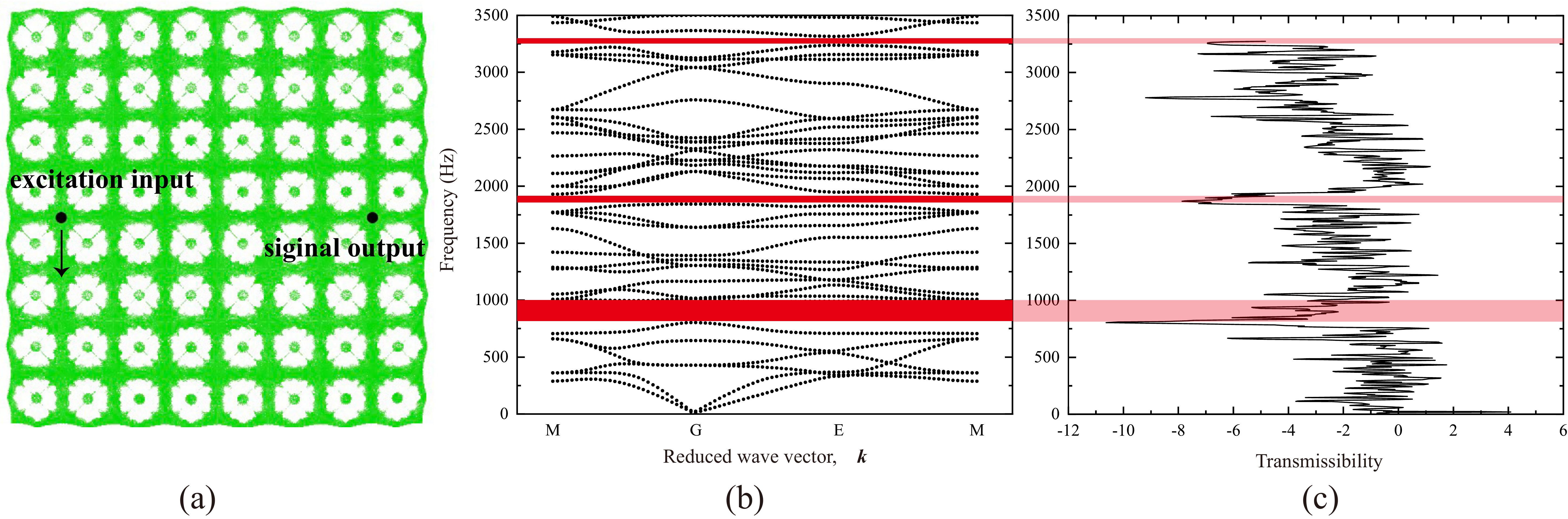}
\caption{(Color online) (a) The deformed model at $\varepsilon=0.14$ for an $8\times8$ structure when the simply supported condition is used. The two black dots characterize the positions for signal input and output, and the arrow highlights the direction of the excitation. (b) Dispersion relations for the deformed state in (a). (c) The corresponding transmission curve.}\label{fig13}
\end{figure*}

To illustrate the validity of the numerically calculated predictions of the bandgaps in Figure \ref{fig12b}, we investigate the steady-state dynamical behavior of an $8\times8$ structure in response to a harmonic excitation. The post-buckling state is first generated when the macroscopic mean strain is identical to $0.14$. Figure \ref{fig13}(a) sketches that we exert a vertical excitation at the left black point and measure the magnitude of the displacement at the right black point. The dynamical analysis is performed by steady-state dynamics step in Abaqus. We define the transmissibility by
$\operatorname{log}(|w_{ou}|/|w_{in}|)$, where $w_{in}$ and $w_{ou}$ denote the displacements of the excitation point and the signal output point, respectively. Figure \ref{fig13}(b) plots the dispersion curves within the range of 0 Hz -3500 Hz while the associated transmissibility is shown in Figure \ref{fig13}(c). It turns out that the transmission is attenuated significantly in the frequency ranges of bandgaps. This indicates the validity of our bandgap prediction as well as the tunability of the wave property of the proposed metamaterial structure through large deformation. Moreover, we also find that transmissibility reduces notably nearly 2500 Hz where no complete bandgap appears. Therefore, in addition to the complete bandgaps indicated by red color, there also exists a directional bandgap in the buckled configuration.  

\begin{figure}[!h]
\centering\includegraphics[width=9cm]{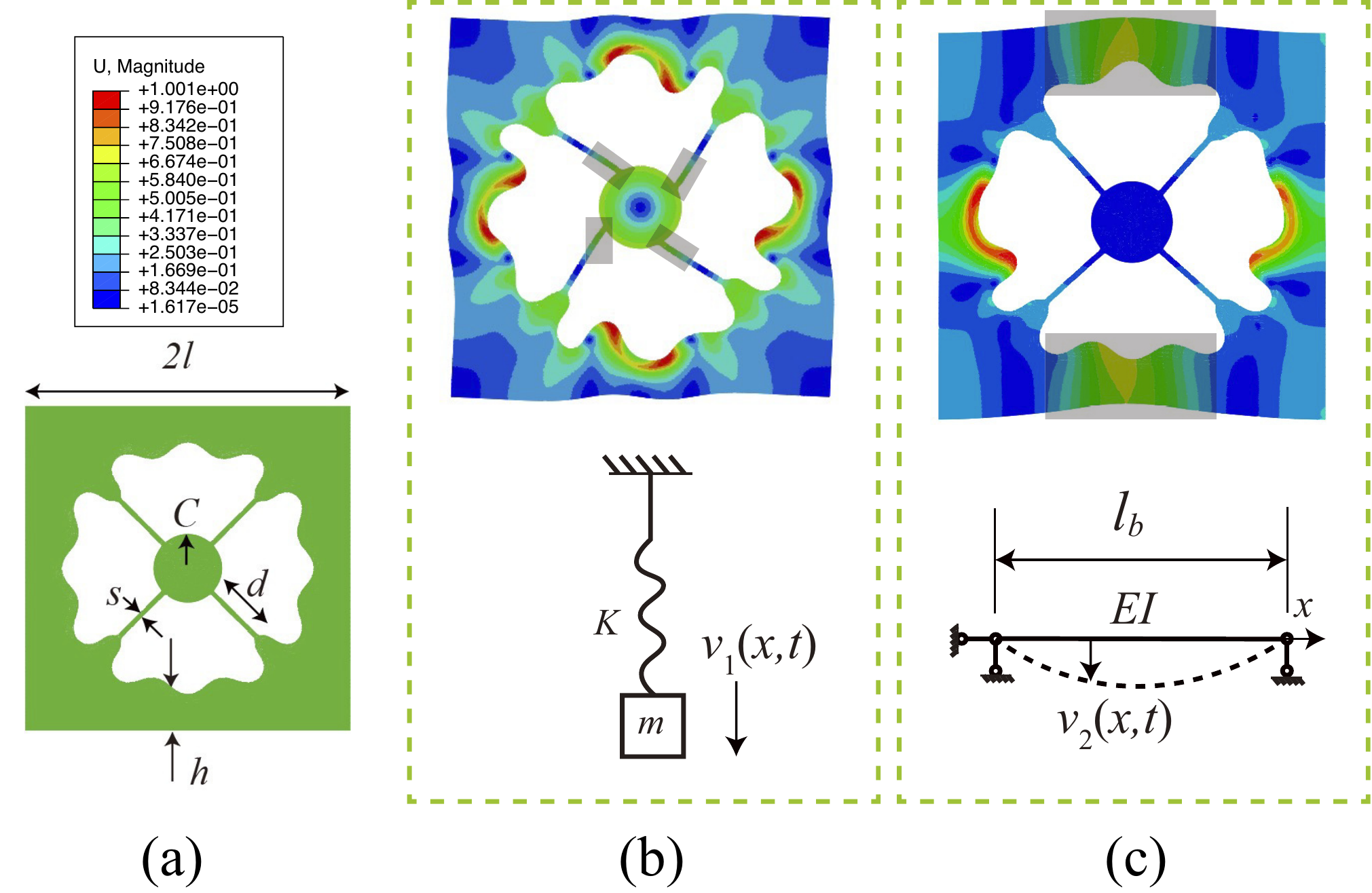}
\caption{(Color online) (a) The post-buckling state when $\varepsilon=0.14$ with several new defined geometrical parameter. (b) Vibration mode of the lower bound point on the bandgap for the point G (the frequency is 801.95Hz). Below is a simplified mass-spring model of the shaded region. (c) Vibration mode of the upper bound point on the bandgap for the point G (the frequency is 1009.3Hz). Below is a simply supported beam model of the shaded region.}\label{fig14}
\end{figure}

Finally, we inspect whether the bandgap in the low-frequency range in Figure \ref{fig12b} is caused by local resonance. For that purpose, we shall focus on the vibration mode of the unit cell associated with periodic boundary conditions when $\varepsilon=0.14$ and attempt to establish a simplified theoretical model. As indicated in Figure \ref{fig14}(a), we have denoted the thickness and length of the beam by $s$ and $d$, respectively, in the post-buckling configuration. In particular, the radius $C$ is preserved as the core is rigid and the height $h$ will be explained later. Meanwhile, we intend to interpret the mechanism resulting in the low-frequency bandgap. Thereby the related vibration modes at the point G for the lower and upper boundaries of the bandgap are explored. The corresponding frequencies are given by 801.95 and 1009.3 Hz, respectively, and the vibration modes are plotted in Figures \ref{fig14}(b) and \ref{fig14}(c). It can be seen from Figure \ref{fig14}(b) that the displacement mainly concentrates on the peaks and valleys of surface coating as well as the metallic core. In addition, the core has an obvious rotation which pulls the four joint beams. However, the situation in Figure \ref{fig14}(c) is entirely dissimilar, and the metallic core is almost static. This demonstrates the existence of a mode transition between the lower and upper bounds of the low-frequency bandgap. According to the mode feature, we assume that the vibration in Figure \ref{fig14}(b) can be described by a mass-spring system comprised of four springs (the dark part) and a rigid body (the metallic core). The vibration equation is then given by
\begin{equation}
m \ddot{v}_1+Kv_1=0,
\end{equation} 
where $v_1$ is the vertical displacement, $m=\rho_{st}\pi C^2$ the mass of the metallic core and $K=16\alpha G(1+\nu) s/d$ the stiffness of spring. Compared to the rigid core made of No. 45 steel, the contributions of the material surrounding the core as well as the spring to the total mass can be neglected. To offset the effect of the displacements in the coating and matrix, we introduce a correction factor $\alpha$ in the spring stiffness. The length of each spring is $d/2$. In doing so, the natural frequency $f$ of the simplified equivalent system takes the following form
\begin{equation}
f=\dfrac{1}{2\pi}\sqrt{\dfrac{K}{m}}=\dfrac{2}{\pi C}\sqrt{\dfrac{G(1+\nu)\alpha s}{\rho_{st}d\pi}}.\label{eq5_4}
\end{equation}
For a specified post-buckling state, only the correction factor $\alpha$ is unknown, and we shall determine the value of $\alpha$ later.

Next, we deal with the vibration pattern in Figure \ref{fig14}(c). It presents a highly localized vibration, which conforms to the fact that the corresponding dispersion curve is practically horizontal (see Figure \ref{fig12b}). We assume that the shaded areas suffering local vibration occupy the same natural frequency. In addition, these two parts can be viewed as simplified supported beams. As a result, it is convenient to take one beam as a representative structure. As the surface coating is very thin and the matrix is nearly incompressible, we approximately take the bending stiffness of the beam as $2G(1+\nu)I$ where $I=l_3h^3/12$ is the moment of inertia of the cross-section and employ the original density of the matrix in the vibration model. Here the length $l_3$ in the third direction of a plane-strain problem can be prescribed by unity without loss of generality. In addition, the average beam height $h$ is defined by $(d_p+d_v)/2$ where $d_p$ depicts the distance from the lateral boundary to a peak in the post-buckling configuration while $d_v$ the counterpart to a valley. Denoting the deflection by $v_2(x,t)=\tilde{v}_2(x)\textrm{e}^{-\textrm{i}\omega t}$ with $\omega$ being the natural angular frequency, we obtain the governing equation 
\begin{equation}
\dfrac{\textrm{d} ^4\tilde{v}_2}{\textrm{d} x^4}-\omega ^2\dfrac{\rho}{EI}\tilde{v}_2=0,\label{eq5_5}
\end{equation} 
where $\rho=\rho_0 h l_3$ represents the mass per unit length. The boundary conditions read 
\begin{equation}
\tilde{v}_2(0)=\dfrac{{\rm d}^2\tilde{v}_2}{{\rm d}x^2}\bigg |_{x=0}=\tilde{v}_2(l_b)=\dfrac{{\rm d}^2\tilde{v}_2}{{\rm d}x^2}\bigg |_{x=l_b}=0, \label{eq5_6}
\end{equation}
where $l_b$ signifies the effective beam length and can be evaluated by seeking the boundary between blue and grey meshes (blue denotes a nearly zero deformation). Solving equation (\ref{eq5_5}) subject to the boundary condition (\ref{eq5_6}) yields
\begin{equation}
f=\frac{\omega}{2\pi}=\frac{\pi h}{2l^2_b}\sqrt{\frac{G(1+\nu)}{6\rho_0}}.\label{beam}
\end{equation}

\begin{figure}[!h]
\centering\includegraphics[scale=0.32]{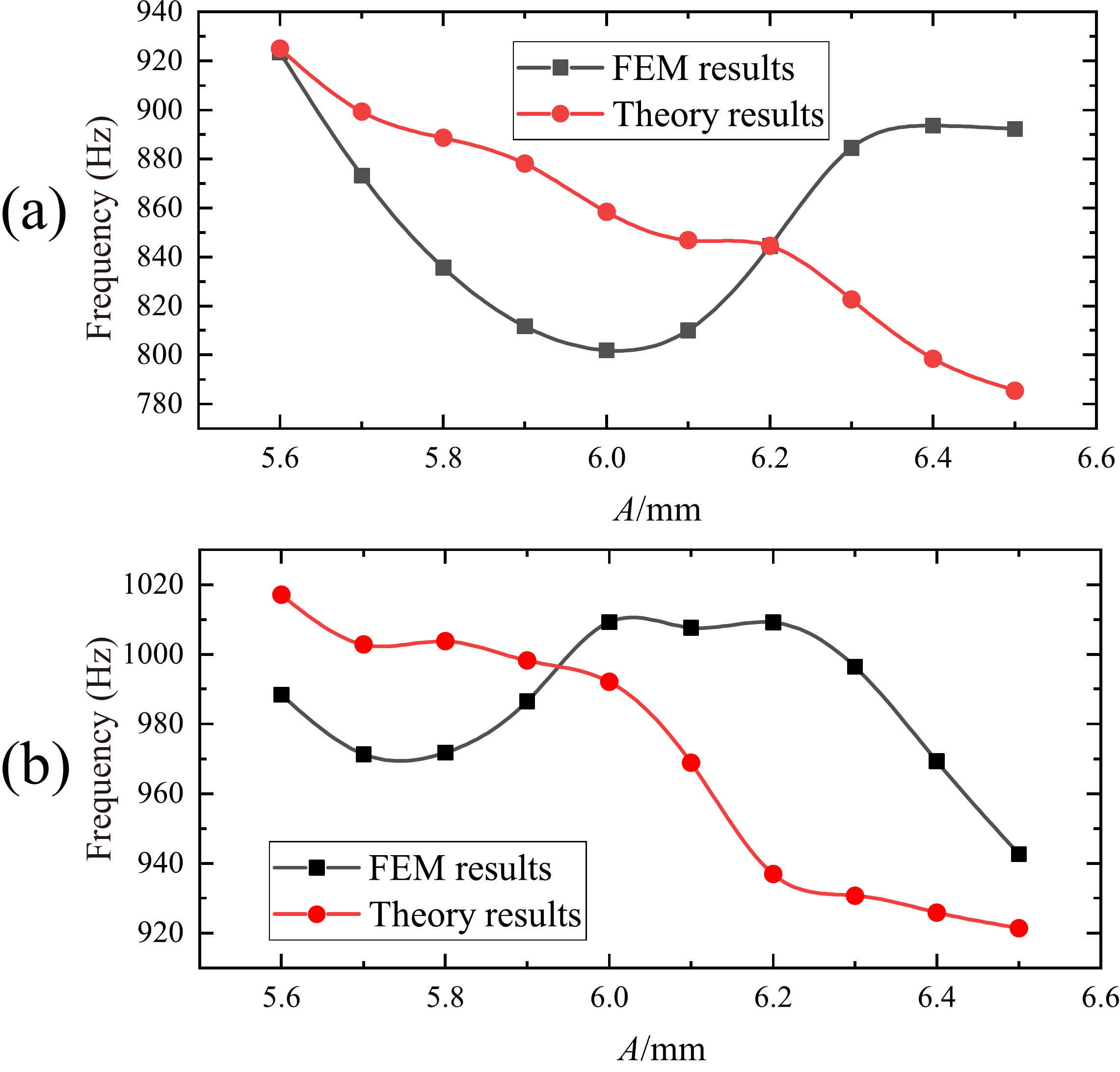}
\caption{(Color online) Comparisons between the FE and theoretical results as the hole radius A is changed. (a) The lower bound natural frequency of the bandgap. (b) The upper bound natural frequency of the bandgap.}\label{fig15}
\end{figure}   

Subsequently, we plan to determine the correction factor $\alpha$ in (\ref{eq5_4}) and to check the validity of the proposed theoretical models. To this end, we fix $T=0.1$ mm from now on and alter the hole radius from $A=5.6$ mm to $A=6.5$ mm. In this case, another radius $B$ will change simultaneously. Besides, all other parameters keep the same as those used in obtaining Figures \ref{fig11} -- \ref{fig13}. We denote the thermal strain of the coating by $\zeta\Delta t$ with $\zeta$ being the thermal expansion coefficient and $\Delta t$ the temperature variation. It is emphasized that the thermal strains are all prescribed by 0.5 in these post-buckling states. Accordingly, the macroscopic mean strain $\varepsilon$ may be different. We calibrate the value of $\alpha$ to available FE results by use of the method of least squares. After a routine calculation, we obtain $\alpha=1.616$. Furthermore, if $A=6$ mm, we get $C$=2 mm, $d$=4.5 mm, $s$=0.31mm, $h$=3.1mm, and $l_b$=8.9 mm. The frequency of lower bound is 858.32 Hz, which is a bit higher than the FE solution $801.95$ Hz while the counterpart of the upper bound is 992.06 Hz which is quite close to the FE solution 1009.3 Hz. The associated comparisons for different $A$ between the theoretical and FE results are shown in Figure \ref{fig15}. One observes the remarkable deviations as the vertical axis varies in a compact range. Actually, the maximum relative errors are nearly $12\%$ and $7\%$ in Figure \ref{fig15}(a) and Figure \ref{fig15}(b), respectively, and the latter justifies the fundamental assumption that natural frequencies of these vibrational parts in Figure \ref{fig14}(c) are quite similar. Furthermore, the beam length $l_b$ has a drastic effect on the frequency. Hence any inaccurate evaluation of $l_b$ may cause a relatively noticeable deviation. In summary, the simply spring and beam models pave a convenient way to estimate the range of a low-frequency bandgap.

\section{Parametric studies on the metamaterial structure with metallic cores}
As indicated in Figure \ref{fig10}, there are many parameters involved in the unit cell of the designed metamaterial structure, i.e. the half length $L$, inner radius $A$, interfacial radius $B$, core radius $C$, beam thickness $S$, and the modulus ratio $\xi$. We assume that the beam thickness $S$ has a minor influence on pattern formation and wave property of the metamaterial so that it is specified by 0.4 mm from now on. In addition, the length $L$ is 10 mm and we introduce a new parameter $\eta$ (normalized thickness ratio) defined by $\eta=(L-B)/(B-A)$ to signify the thickness ratio between the matrix and the pore coating. As a result, we are concerned with the effects of the remaining parameters on the onset of surface buckling as well as the band structures.

\subsection{Onset of surface wrinkling}
We first discuss the buckling response. The macroscopic mean strain $\varepsilon$ is taken as the bifurcation parameter in a consistent manner. We are focusing on four free parameters $A$, $\eta$, $C$, and $\xi$, and a two-dimensional picture is a better way to depict how a specific parameter affects the buckling initiation. Hence we shall fix three of these parameters and then exhibit the dependence of the critical strain $\varepsilon_{cr}$ on the left one. As the inner surface is not entirely traction-free, the hoop stress distribution in the circumferential direction is inhomogeneous at the beginning of the deformation in spite of the fact that there is no surface wrinkle. Accordingly, we define the bifurcation threshold as follows. By tracking the evolution of the hoop stress at the free surface, we take where a sinusoidal inhomogeneity of the circumferential stress is observed in FE calculations as the critical strain $\varepsilon_{cr}$. In addition, the other parameters are given by $\mu=1.08$ MPa, $\kappa=2$ GPa, $\rho_0=1050$ $\textrm{kg}/\textrm{m}^3$, $\nu_{st}=0.269$, $E_{st}=206$ GPa and $\rho_{st}=7890 $kg$/\textrm{m}^3$.

\begin{figure}[!h] 
\centering\includegraphics[scale=0.42]{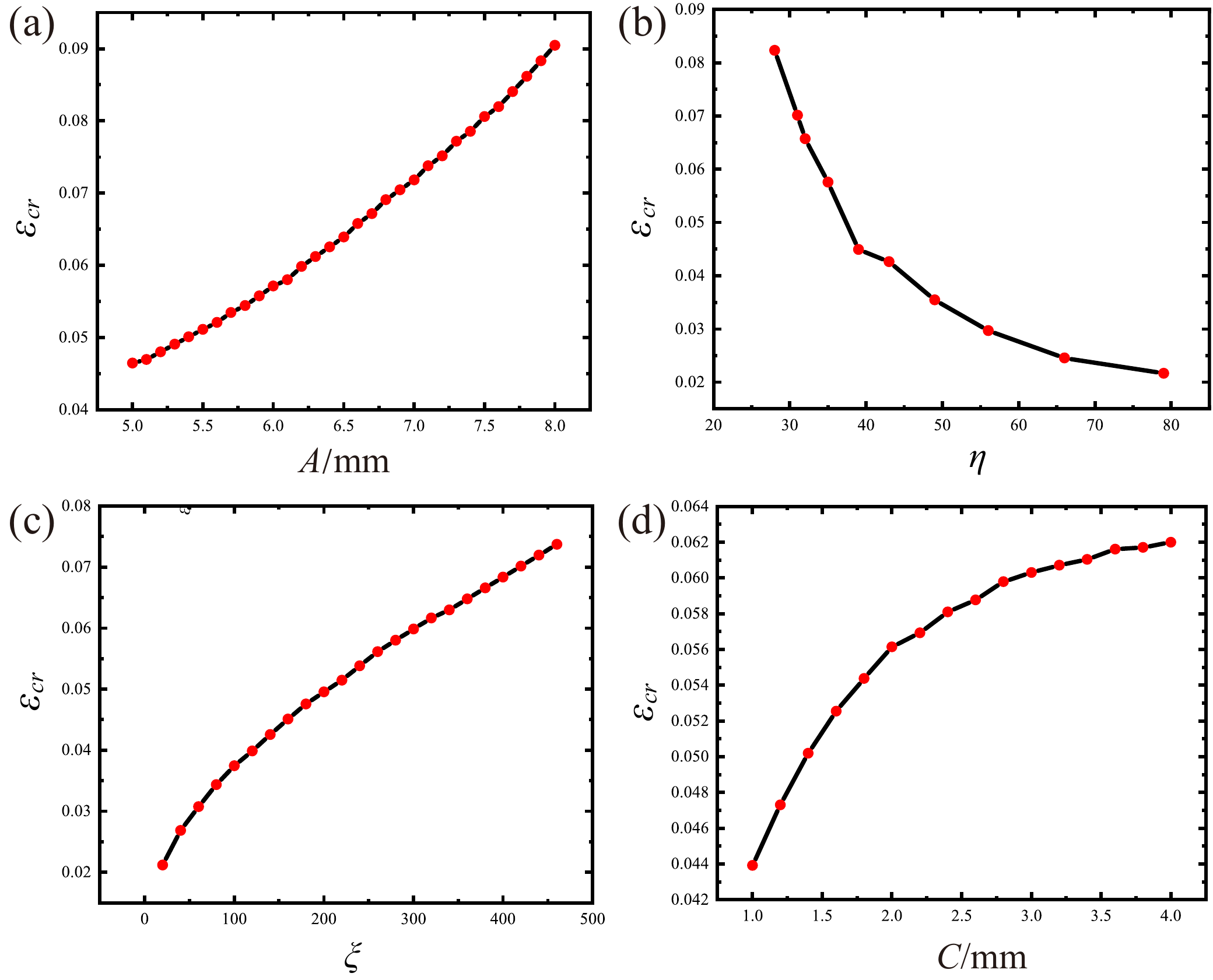}
\caption{(Color online) The critical strain as functions of the (a) inner radius $A$, (b) scaled thickness ratio $\eta$, (c) the modulus ratio $\xi$, and (d) the core radius $C$ based on FE simulations.}\label{fig16}
\end{figure} 

We investigate how the unit cell behaves with varied parameters and refer to Section 2 for the meshing procedure. Specifying $\xi=300$, $C=2$ mm, and $T=0.11$ mm where $T=B-A$ defined in Figure \ref{fig1} is the thickness of the pore coating, we plot the critical strain $\varepsilon_{cr}$ versus the radius of the center hole $A$ in Figure \ref{fig16}(a). Actually, we only obtain the values of $\varepsilon_{cr}$ for the red dots and all the continuous curves in Figure \ref{fig16} are obtained by joining the dots together. It can be seen that the critical strain is an increasing function as $A$ varies from 5 mm to 8 mm. Specifically, it is nearly a straight line. It can be concluded that a lower porosity prones to destabilize the structure. Figure \ref{fig16}(b) shows the dependence of $\varepsilon_{cr}$ on the thickness ratio $\eta$ when $\xi=300$, $C=2$ mm, and $A=6$ mm. In this case, altering $\eta$ is equivalent to tuning $B$ and a greater $\eta$ implies a larger volume ratio of the soft matrix. It is found that the structure is less stable if we enlarge the volume of the soft matrix. Next, we take $A=6.2$ mm, $C=2$ mm, and $T=0.11$ mm (or equivalently $B=6.31$ mm) and plot the relation between $\varepsilon_{cr}$ and the modulus ratio $\xi$ in Figure \ref{fig16}(c). In contrast with the existing results for film/substrate structures, either planer or curved systems, where a stiffer coating (or film) always reduces the critical strain for surface wrinkling \citep{jlc2018}, the larger the stiffness mismatch between the surface coating and the matrix is, the higher the critical strain attains in periodically porous elastomers. Note that $\varepsilon$ only denotes the macroscopic mean strain. We have carefully checked the thermal strain (true strain) recorded in Abaqus as well and find the same tendency as that in Figure \ref{fig16}(c). In other words, a stiffer coating can retard surface wrinkling. To seek a possible factor causing such an anomalous consequence, we replace the periodical boundary conditions by the fixed counterparts for the $1\times1$ unit cell in the heating process. In this way, the mean strain $\varepsilon$ can not be used to measure the volume increase. We therefore adopt the thermal strain as the bifurcation parameter and find that the critical strain is a decreasing function of the modulus ratio $\xi$. This indicates the vital role of the boundary conditions in destabilizing the structure. Finally, we let $A=6.2$ mm, $T=0.11$ mm, $\xi=260$ and  regard the radius of the metallic core (cylindrical oscillator) $C$ as the free parameter. Figure \ref{fig16}(d) displays the curve of $\varepsilon_{cr}$ versus $C$. It can be seen that a larger core incurs a more stable structure.

\begin{figure}[!h] 
\centering\includegraphics[scale=0.4]{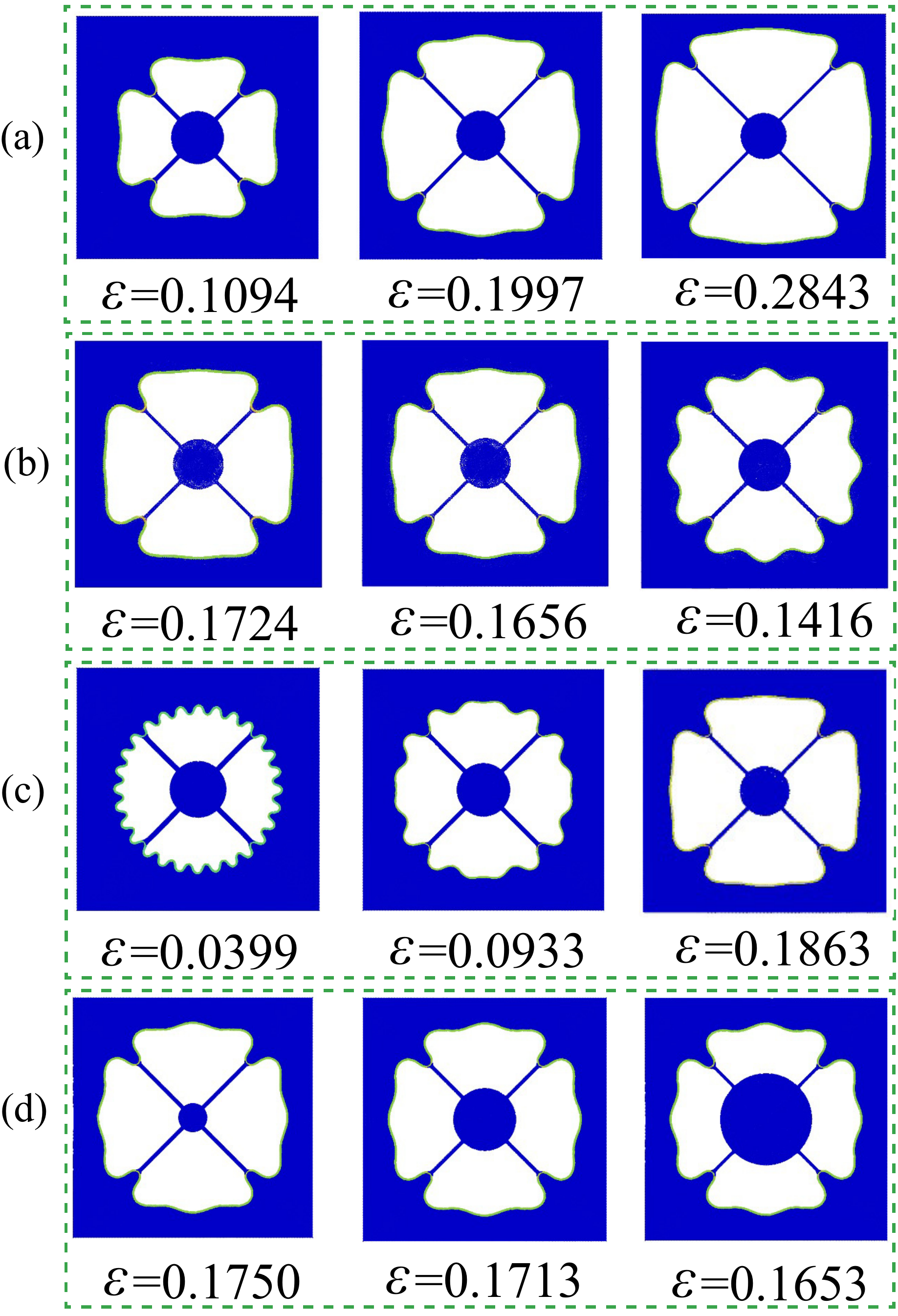}
\centering
\caption{(Color online) Selected post-buckling configurations for various parameters. The value of macroscopic mean strain $\varepsilon$ is given below each subfigure. (a) The radius of the hole $A$ is given by 5 mm, 6.5 mm, and 7.5 mm from the left to the right. (b) The thickness ratio $\eta$ is given by 97/3, 389/11, and 39. (c) The modulus ratio $\xi$ is given by 20, 160, and 500. (d) The core radius $C$ is given by 1 mm, 2.6 mm, and 4 mm.} \label{fig17}
\end{figure} 

Figure \ref{fig17} sketches several post-buckling patterns associated with the parametric study in Figure \ref{fig16}. For instance, only the hole radius $A$ is varied in Figure \ref{fig17}(a) and all other parameters are identical to those used in obtaining Figure \ref{fig16}(a). Furthermore, the corresponding strain value is provided below each figure. It is found that more wrinkles can be generated as the hole radius grows. Seen from Figure \ref{fig17}(b) where $\eta$ is the free parameter, all three buckling patterns share the same wavenumber, and a larger eta incurs an obvious modulation. Figure \ref{fig17}(c) implies that a greater stiffness mismatch produces a lower wavenumber. Yet the buckling pattern is almost independent of the core radius $C$, as shown in Figure \ref{fig17}(d).

\subsection{Bandgaps}
In this subsection, we intend to conduct an exhaustive parametric analysis on the band structures of the designed metamaterials with local oscillator. This is necessary to unravel the influence of distinct parameters on wave propagation and further to help design the metamaterial structure for a particular application. The fixed parameters have been written in the front of this section, and the unit cell ($1\times1$ structure) is used.

\begin{figure*}[!h] 
\centering\includegraphics[width=18cm]{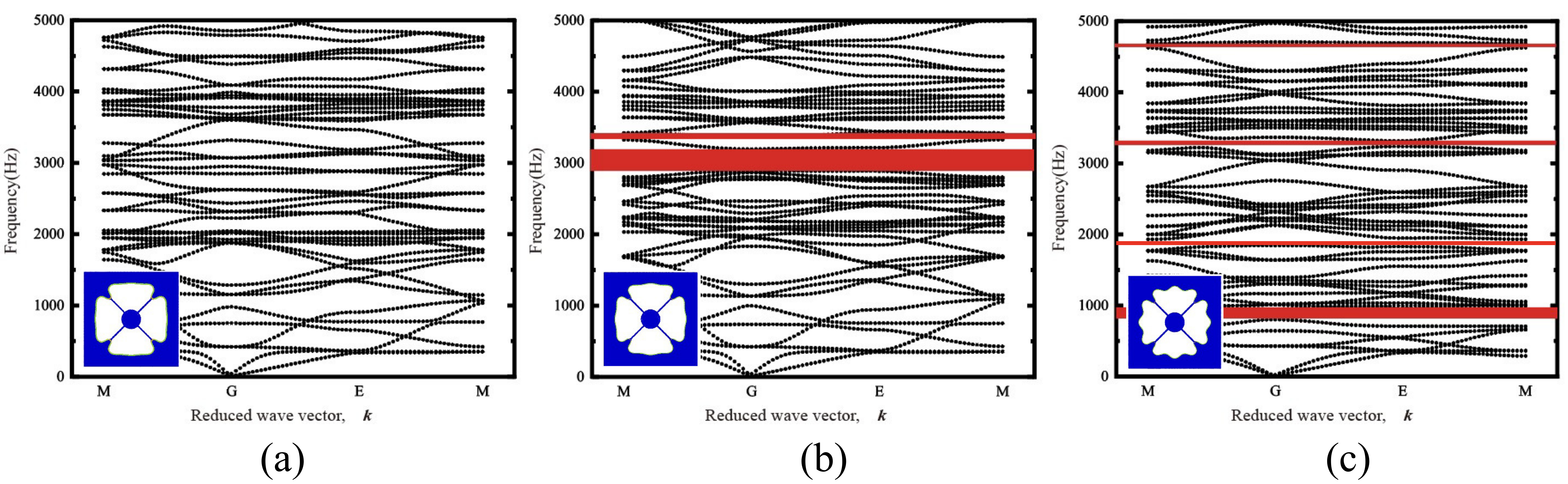}
\centering
\caption{(Color online) Dispersion relations of different values of ($\eta$, $\varepsilon$) which are given by (97/3, 0.1724), (389/11, 0.1656), and (39, 0.14), respectively, from left to right. The red areas express the bandgaps.}\label{fig18}
\end{figure*} 

It is pointed out that the stiff coating is usually thin so as to engender surface wrinkling and further to yield more wrinkles.  In our another separate study regarding experimental observation of surface wrinkles induced by swelling, it is found that fabrication of a sample may bring imperfection in the coating thickness \citep{ljc2022} and may further lead to a relatively notable deviation of the surface pattern compared with the theoretical prediction. In other words, it is not an appropriate way to regulate bandgaps by the coating thickness $T$, or equivalently, the parameter $\eta$. On the one hand, inspired by the results in Figures \ref{fig9} and \ref{fig12}, we expect that bandgaps may be set off by a specific buckling pattern. Hence we fix $A=6$ mm, $C=2$ mm, $\xi=300$ and consider that the coating thickness $T$ ranges from 0.1mm to 0.2 mm in a step increment 0.01 mm. It can be readily checked that the associated $\eta$ lies in $(19,39)$. On the other hand, previous studies in soft metamaterials have revealed that buckling instability is able to induce a wider bandgap. We then perform FE simulations to track buckling evolution and take a post-buckling state as the base state for wave propagation (with a relatively larger $\varepsilon$). Especially, the same maximum load (thermal strain $\zeta\Delta t=0.5$) is applied in all calculations. It is found from our FE simulations that the chosen buckled state can not filter any wave until $T$ is equal to or less than $0.11$, or equivalently, $\eta\geqslant 389/11$. We exhibit three dispersion relations in Figure \ref{fig18}. The associated pairs of $(\eta, \varepsilon)$ are given in the figure caption. When $\eta=97/3$ (or $T=0.02$ mm), there is no bandgap. As $\eta$ increases up to $389/11$, we observe two bandgaps including a wider one in the range of 2887.3 Hz-3193.7 Hz (gap width 306.4 HZ) and a narrow one whose bandgap range is 3324.6 Hz-3364.2 Hz (gap width 39.6 Hz). We mention that, when $\eta=39$ and $\varepsilon=0.14$, Figure \ref{fig18}(c) is identical to Figure \ref{fig12b}. There are four bandgaps and the widest one ranges from 801.95 Hz to 984.24 Hz with a width 182.29 Hz.

\begin{figure}[!h] 
\centering\includegraphics[width=9cm]{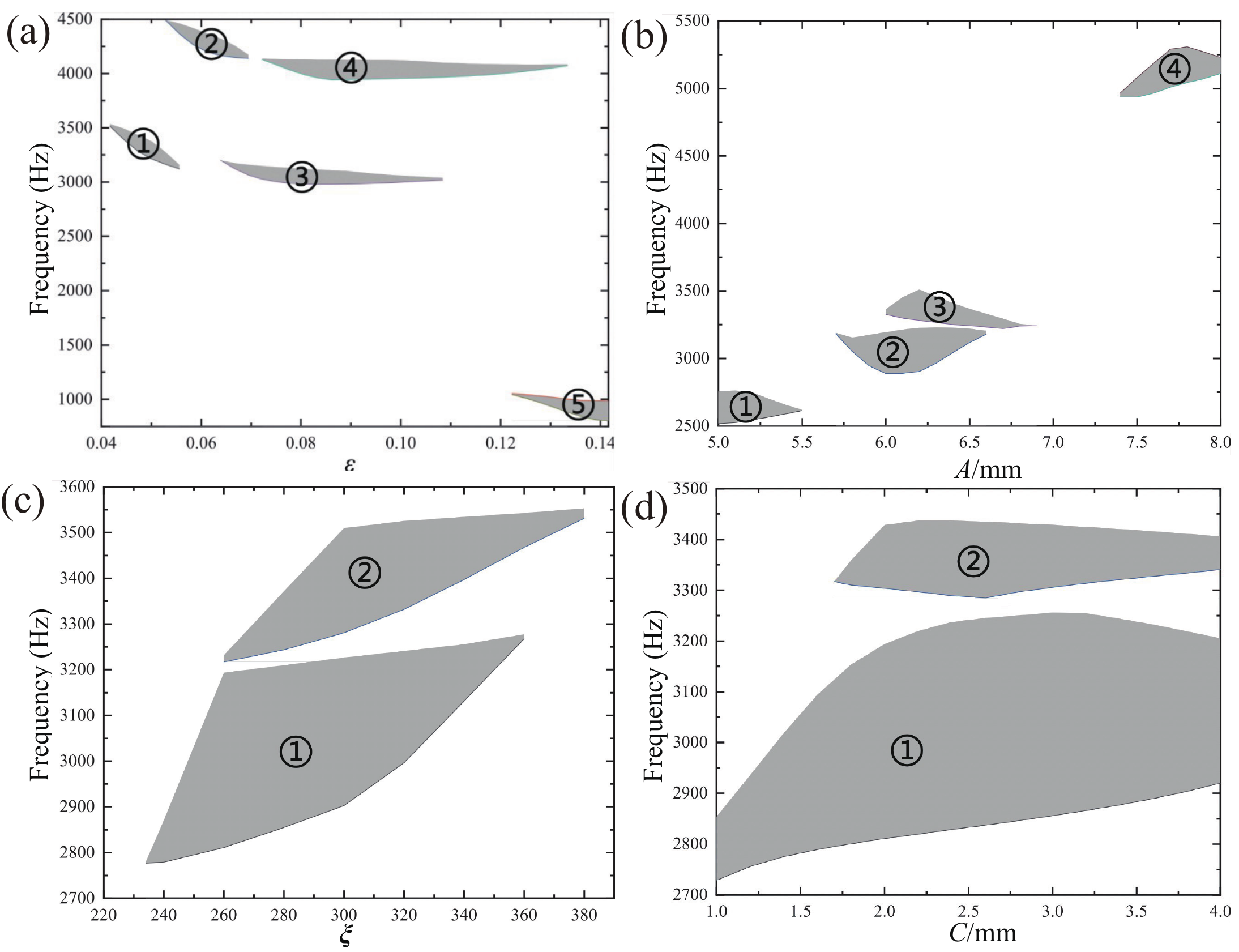}
\centering
\caption{(Color online) Effects of (a) the applied mean strain $\varepsilon$, (b) the radius of the center cavity, (c) the modulus ratio between surface coating and matrix, and (d) the radius of the local resonance on the bandgaps which have been highlighted by grey color.} \label{fig19}
\end{figure} 

Given $A=6$ mm, $C=2$ mm, $B=6.1$ mm and $\xi=300$, we summarize the evolution of bandgaps as a function of the applied macroscopic mean strain $\varepsilon$ in Figure \ref{fig19}(a). There are in total five bandgaps in the frequency range 0 Hz - 4500 Hz. No bandgap is found in the undeformed state and at the beginning of the deformation. As $\varepsilon$ exceeds 0.0417, a bandgap ranging from 3210.9 Hz-3376.5 Hz (width 165.6 Hz) occurs, labelled as No.1. With increasing $\varepsilon$, an additional bandgap No.2 lying in 4200.5 Hz-4370.1 Hz (width 169.6 Hz) appears, and both the upper and lower bounds of bandgaps No.1 and No.2 decline until these two bandgaps are closed. Yet bandgaps No.3 (2987.7 Hz-3129.4 Hz, width 141.7 Hz) and No.4 (3942.1 Hz-4128.2 Hz, width 186.1 Hz) can be generated when $\varepsilon$ enlarges continuously. A low-frequency bandgap (801.9 Hz-984.2 Hz, width 182.3 Hz) is triggered if $\varepsilon$ exceeds 0.12. Later, our FE simulations reach the maximum applied load and thereby we stop at around $\varepsilon=0.142$. Actually, Figure \ref{fig19}(a) unravels how the band structure evolves relative to the surface patterns in Figure \ref{fig11}. Linking with the analysis of the low-frequency bandgap in the previous section, it can be concluded that highly localized surface pattern opens the bandgap in high frequency ranges and local resonance induces the low-frequency bandgap. 

Figure \ref{fig19}(b) plots the band structure versus the cavity radius $A$ with $T=B-A=0.11$ mm, $C=2$ mm and $\xi=300$. The thermal strain also attains 0.5 in the coating for consistence (see also Figures \ref{fig19}(c) and \ref{fig19}(d)). In fact, these  parameters have been employed in plotting Figure \ref{fig18}(b). In this situation, we do not find a low-frequency bandgap. Seen from Figures \ref{fig12b} and \ref{fig15} that low-frequency bandgap emerges if $T=0.1$ mm. This manifests that the coating thickness is a vital parameter in regulating low-frequency bandgaps which can be estimated by two simplified theoretical models presented in the previous section. Notwithstanding, the bandgap in Figure \ref{fig18}(b) where $T=0.11$ mm is wider such that we are also interested in the case. Four bandgaps are obtained as $A$ increases from 5 mm to 8 mm. In general, a greater hole may produce a bandgap with high frequency. For the bandgap No.1, with growing $A$, the lower boundary of the bandgap rises slowly whereas the upper boundary increases at first and decreases afterwards. The widest frequency range of this bandgap is 2514.8 Hz-2754.4 Hz (bandgap width 239.6 Hz) and it occurs when $A\approx$ 5 mm. For the bandgap No.2, the upper boundary keeps nearly horizontal yet the lower boundary declines sharply and then increases. The widest frequency range of this bandgap is 2888.5 Hz-3212.6 Hz (bandgap width 324.1 Hz) and the accompanied $A$ is around 6.1 mm. If $A$ is between 6 mm and 6.5 mm, another bandgap No.3 can be seen which is quite close to bandgap No.2. We mention that the proposed metamaterial structure cannot prevent any waves if $A$ locates at $(5.5, 5.54)$ (mm) and $(7, 7.4)$  (mm). The widest frequency ranges read 3280.6 Hz-3509.6 Hz and 5009.7 Hz-5291.7 Hz, respectively, for the bandgaps No.3 and No.4, and the associated cavity radius are approximately given by 6.2 mm and 7.7 mm.

It is noteworthy that the widest frequency range in Figure \ref{fig19}(b) is attained at $A\approx$ 6.2 mm where two separated bandgaps can be incurred. Moreover, such a hole radius would not cause a bent and thin lateral boundary which will result in narrow bandgaps. As a result, we choose $A=$ 6.2 mm to continue the subsequent calculations.      

Figure \ref{fig19}(c) displays the bandgaps in terms of the modulus ratio $\xi$. The parameters are given by $A=6.2$ mm, $B=6.31$ mm ($T=0.11$ mm) and $C=2$ mm. In particular, we take $\xi\in[180,460]$ for ensuring the convergence of our calculations in Abaqus with respect to the $1\times1$ unit cell with the same maximum applied thermal strain, i.e. 0.5. It is found from our calculations that two bandgaps are found at the frequency range 2700 Hz-3600 Hz. The lower bound of each bandgap area is an increasing function of $\xi$. In each bandgap area, the bandgap width increases first and then decreases. The widest frequency range of bandgap No.1, attained at $\xi\approx260$, is 2810.8 Hz-3193.8 Hz (width 383 Hz) while the counterpart of bandgap No.2 is obtained at $\xi\approx300$ and is given by 3280.6 Hz-3509.6 Hz (width 229 Hz). Furthermore, we fix  $A=6.2$ mm, $B=6.31$ mm ($T=0.11$ mm), $\xi=260$ and show the bandgap distribution in Figure \ref{fig19}(d). Similarly, there are also two bandgaps in $C\in[1,4]$ (mm). It can be found that the widest frequency ranges of these two bandgap areas are 2828.4 Hz-3237.2 Hz (width 408.8 Hz) and 3284.9 Hz-3435.2 Hz (width 150.3 Hz), respectively. In addition, the widest bandgaps can be obtained at $C\approx2.4$ mm and $C\approx$ 2.6 mm, respectively.

Currently, we have completed the parametric studies on both the buckling threshold and the bandgap evolution of the designed metamaterial with metallic cores. In particular, we have validated these predicted bandgaps by the associated transmission curves similar to Figure \ref{fig13}. Yet these validations are omitted in order to save length. It is worth mentioning that there are more parameters involved compared to the periodically porous elastomers without coatings. This feature makes the new structure more tunable and makes the problem more complicate as well. As a result, the current study only reveals some fundamental characteristics on the buckling, post-buckling and wave property of the proposed metamaterial. We anticipate that the results shown in Figures \ref{fig16}-\ref{fig19} are helpful in selecting a specific parameter for a certain purpose in practical applications.

\section{Concluding remarks}
Inspired by pattern formation in tubular structures subject to growth, swelling, and internal pressure \citep{mg2011,bec2015,dll2019,cnn2021}, we designed a new class of tunable and switchable metamaterial structure by imposing a thin coating on each cavity of a periodically porous elastomer. The main advantage is that finite deformation and surface instability can be generated by a non-contact loading scenario. In particular, the unit cell contains only one pore and we adopt the $1\times1$ structure to represent it. We have taken the thermal loading as a paradigm and numerically explored the deformation and pattern formation in the unit cell associated with periodic boundary conditions in the commercial software Abaqus. We considered that the coating is much stiffer than the matrix and the thermal strain can merely be induced within the coating. In all illustrative calculations, both materials are assumed to be modeled by nearly incompressible neo-Hookean constitution. The periodic boundary conditions and Bloch boundary conditions have been implemented by writing Python scripts, the validity of which was confirmed by reproducing some existing results. It is found that the unit cell will expand driven by heating-induced deformation in the coating and the hole becomes larger and larger. To clearly depict how big the unit cell will be in a deformed state, we have employed the macroscopic mean strain $\varepsilon$ as the bifurcation parameter and found that surface wrinkles appear at a critical value of $\varepsilon$. We then studied the bandgap in the proposed metamaterial by FE analysis in Abaqus. It turns out that several narrow bandgaps in high-frequency ranges can be unfolded by surface wrinkling. This implies the possibility of using such a structure to prevent wave propagation with specific frequencies.

We further refined the metamaterial by adding metallic cores to the center of each cavity. It is expected that these cores can produce local resonance which may give rise to a low-frequency bandgap. It turns out that the surface wrinkling can also be observed at a critical strain. Furthermore, a low-frequency bandgap was triggered when geometrical and material parameters belong to a certain domain. In order to reveal the mechanism behind the low-frequency bandgap, we developed two simplified models containing a mass-spring system and a beam system, from which the lower and upper bounds of a low-frequency bandgap can be computed. Finally, a parametric study was performed for the onset of surface wrinkling and bandgaps of the refined structure. In doing so, the effects of different geometrical and material parameters on the surface instability and wave property were clarified. A remarkable finding is that the structure is more stable if the modulus ratio $\xi$ is greater. This contradicts the available results concerning surface wrinkling of tubular structures which are always susceptible to surface wrinkling for a larger modulus ratio \citep{lcf2011,jlc2018}. Finally, some phase diagrams of bandgaps were shown, which could provide guidance on parameter selection in practical applications. It is expected that the current study will create a new connection between pattern formation in soft materials and wave regulation in metamaterials. The current work is a just preliminary investigation on utilizing surface wrinkles to tune wave propagations. It has been shown that magnetic actuation can also serve as a desirable way to alter material properties or to induce finite deformation such that wave propagations can be controlled \citep{cj2000,yl2018,xtw2018,pwc2020,mwk2021,gmm2022}. Consequently, other contactless loading approach as well as the dynamical behaviors of liquid crystal elastomers will be investigated in the future based on some reduced models \citep{lmd2020,lmd2021}. 

Finally, only numerical analysis was performed in the current work. In recent years, many interesting experimental investigations were conducted with regard to wave propagation in soft porous elastomers. Gao et al. \citep{glb2019} fabricated a porous elastomer made of silicon rubber and tested wave propagation in a buckled configuration induced by compression. The white noise was generated by an electrodynamic shaker and the input and output signals were recorded by accelerometers. In doing so, the predicted bandgap can be experimentally validated. Zhou el al. \citep{lwc2019} further investigated the effect of hard inclusions in a similar setup. These works definitely offer useful guidelines for our future experimental study. It is pointed out that, referring to Liu et al. \cite{ljc2022}, a mold for fabricating the metamaterial proposed in Section 2 has been designed. In addition, we could follow Wang et al. \citep{wcs2014} to design a proper mold for fabricating a porous elastomer with metallic cores. Some related experimental works are underway.

\section*{ Acknowledgments}
This work was supported by the National Natural Science Foundation of China (Project Nos: 12072227, 11991031, 12072225, and 12021002). The Abaqus simulations were carried out on TianHe-1 (A) at the National Supercomputer Center in Tianjin, China.
\bigskip

\bibliography{refer}

\end{document}